\documentclass[reprint,twocolumn,superscriptaddress,showpacs,nofootinbib,notitlepage,preprintnumbers,aps,prd,floatfix]{revtex4-2}
\usepackage[utf8]{inputenc}
\usepackage{graphicx}
\usepackage{latexsym,amsmath,amssymb,lmodern,float,url}
\usepackage{qcircuit}
\usepackage{natbib}
\usepackage{color}
\usepackage{microtype}
\usepackage{bbold}
\usepackage{slashed}
\usepackage{multirow}
\usepackage{tikz}
\usepackage{xcolor}
\usepackage{mathrsfs}  
\usetikzlibrary{shapes}
\usepackage{adjustbox}
\usepackage{todonotes}
\usepackage{braket}
\usepackage[plain]{fancyref}
\usepackage[normalem]{ulem}

\newcommand{\fig}[1]{Fig.~(\ref{#1})}

\newcommand{\tref}[1]{Table~\ref{#1}}
\newcommand{\eref}[1]{Eq.~(\ref{#1})}

\usepackage[colorlinks=true,backref=false, linktocpage=true,
citecolor=blue,urlcolor=blue,linkcolor=blue,pdfpagemode=UseOutlines]{hyperref}
\usepackage{cleveref}

\hypersetup{%
 bookmarksnumbered=true,
 pdftitle = {},
 pdfsubject = {},
 pdfauthor = {},
 pdfkeywords = {}
}

\newcommand{\btt}{\mathbb{BT}}

\newcommand{\sdg}{S+D}

\DeclareMathOperator{\tr}{Tr}

\renewcommand{\d}{\mathrm d}

\definecolor{vermillion}{RGB}{213,94,0}
\definecolor{cblue}{RGB}{0,114,178}
\definecolor{corange}{RGB}{230,159,0}
\definecolor{blgreen}{RGB}{0,158,115}
\definecolor{repurple}{RGB}{204,121,167}

\definecolor{mycolor}{RGB}{0,255,0}

\begin{document}
 \preprint{FERMILAB-PUB-24-0612-SQMS-T}
 
\title{Qudit Gate Decomposition Dependence for Lattice Gauge Theories}
\author{Do\~ga Murat K\"urk\c{c}\"uo\~glu}
\email{dogak@fnal.gov}
\affiliation{Superconducting and Quantum Materials System Center (SQMS), Batavia, Illinois, 60510, USA.}
\affiliation{Fermi National Accelerator Laboratory, Batavia, Illinois, 60510, USA}
\author{Henry Lamm}
\email{hlamm@fnal.gov}
\affiliation{Superconducting and Quantum Materials System Center (SQMS), Batavia, Illinois, 60510, USA.}
\affiliation{Fermi National Accelerator Laboratory, Batavia, Illinois, 60510, USA}
\author{Andrea Maestri}
\email{andreamaestrimaestri@gmail.com}
\affiliation{University of Amsterdam, 1090 GL Amsterdam, The Netherlands}
\date{\today}
%\listoftodos
%%%%%%%%%%%%%%%%%%%%%%%%%%%%%%%%%%%%%%%%%%%%%%%%%%%%%%%
\begin{abstract}
 In this work, we investigate the effect of decomposition basis on primitive qudit gates on superconducting radio-frequency cavity-based quantum computers with applications to lattice gauge theory. Three approaches are tested: SNAP \& Displacement gates, ECD \& single-qubit rotations $R(\theta,\phi)$, and optimal pulse control. For all three decompositions, implementing the necessary sequence of rotations concurrently rather then sequentially can reduce the primitive gate run time.
The number of blocks required for the faster ECD \& $R_p(\theta)$ is found to scale $\mathcal{O}(d^2)$, while slower SNAP \& Displacement set scales at worst $\mathcal{O}(d)$.  For qudits with $d<10$, the resulting gate times for the decompositions is similar, but strongly-dependent on experimental design choices. Optimal control can outperforms both decompositions for small $d$ by a factor of 2-12 at the cost of higher classical resources.  
Lastly, we find that SNAP \& Displacement are slightly more robust to a simplified noise model. 
\end{abstract}
%%%%%%%%%%%%%%%%%%%%%%%%%%%%%%%%%%%%%%%%%%%%%%%%%%
\maketitle

\section{Introduction}
While the advent of quantum computers offers the opportunity to investigate new questions in lattice gauge theories (LGT)~\cite{Klco:2021lap,Banuls:2019bmf,Bauer:2022hpo,DiMeglio:2023nsa}, at present the gate resources are prohibitively large~\cite{Gustafson:2024kym,Rhodes:2024zbr}.  Due to the large local Hilbert space of gauge theories, $d$-dimensional qudit-based computers can reduce the resources required~\cite{Gustafson:2021jtq,Gustafson:2021qbt,Popov:2023xft,Kurkcuoglu:2021dnw,Calajo:2024qrc,Gonzalez-Cuadra:2022hxt, Illa:2024kmf, Zache:2023cfj,Gustafson:2024kym}. The potential algorithmic advantage of qudit platforms arises from the increased effective connectivity as native single-qudit $SU(d)$ rotations replace otherwise require non-local multi-qubit circuits~\cite{jankovic2024noisy,Nikolaeva:2021rhq,Mansky:2022bai}. In practice, this may allow for lower individual gate fidelities for the same algorithmic fidelity.  Today, many promising systems for qudits are being explored including trapped ions~\cite{Ringbauer:2021lhi,Nikolaeva:2024wxl,Low:2023dlg}, transmons~\cite{Goss:2022bqd,Luo:2022pxs,Goss:2023frd,Cao:2023ekj,Seifert:2023ous,Wang:2024xbz}, Rydberg arrays~\cite{Kruckenhauser:2022qfi,Cohen:2021axm}, photonic circuits~\cite{Chi:2022uql}, ultra-cold atomic mixture~\cite{Kasper:2020mun}, and superconducting radio frequency (SRF) cavities~\cite{Roy:2024uro}. In this work, we focus on 3D superconducting radio-frequency (SRF) cavities derived from accelerator physics which possess decoherence times on the order of milliseconds~\cite{Roy:2024uro}.  Similar to qubits the possible gate sets are numerous; both for native gates~\cite{krastanov_snap, krastanov2, Kudra_2022,Eickbusch:2021uod} and fault-tolerant computation~\cite{Brennen:2006adc,Howard:2012wif,Bocharov:2018zfk,Yeh:2022eow,Nakajima:2009cxx,Prakash:2021axb,Kalra:2023llu,Booth:2022isz,Glaudell:2022vvu,Glaudell:2024tvz,Prakash:2018rrp}. 

Quantum algorithms for lattice gauge theories generally require a set of fundamental group theoretic operations~\cite{Lamm:2019bik}. The identification of primitive subroutines divides the problem of formulating quantum algorithms for LGT into deriving said group-dependent primitives~\cite{Alam:2021uuq,Gustafson:2022xdt,Zache:2023dko,Gustafson:2023kvd,Gustafson:2024kym,Lamm:2024jnl,Murairi:2024xpc} and group-independent algorithmic design~\cite{Cohen:2021imf,Carena:2022kpg,Gustafson:2022hjf,Zache:2023cfj,Zache:2023dko,Gustafson:2023aai,Carena:2024dzu}. In addition to the group dependence, the primitive subroutines vary based on the digitization of the gauge degrees of freedom.  
Some digitize in the representation basis with a maximum representation encoded~\cite{Zohar:2012xf,Zohar:2012ay,Zohar:2013zla,Zohar:2015hwa,Bazavov:2015kka, Zhang:2018ufj, Unmuth-Yockey:2018ugm, Klco:2019evd, Farrell:2023fgd, Farrell:2024fit, Illa:2024kmf,Ciavarella:2021nmj,PhysRevD.99.114507,Buser:2020uzs,Bhattacharya:2020gpm,Kavaki:2024ijd,Calajo:2024qrc,Murairi:2022zdg}. Alternatively, the $q$-deformed formulation obtains a finite dimensional Hilbert space by changing the symmetry group to be a so-called quantum group~\cite{Zache:2023dko,Zache:2023cfj}.  Other formulations consider truncations within the bases of gauge-invariant states~\cite{davoudi2024scattering,Raychowdhury:2018osk,Kadam:2023gli,Davoudi:2020yln,Mathew:2022nep,Bauer:2021gek,Grabowska:2022uos,Grabowska:2024emw,Li:2024ide}.  Further methods begin with different formulations or perform different approximations exist such as light-front quantization~\cite{Kreshchuk:2020dla,Kreshchuk:2020aiq,Kreshchuk:2020kcz}, conformal truncation~\cite{Liu:2020eoa}, strong-coupling and large-$N_c$ expansions~\cite{Fromm:2023bit,Ciavarella:2024fzw}. Another approach is to formulate an inherently finite-dimensional Hilbert space theory with continuous local gauge symmetry which is in the same universality class as the original theory.  Some methods for this include fuzzy gauge theories which use non-commutative geometry~\cite{Alexandru:2023qzd}, and quantum link models which use rishons and an ancillary dimension~\cite{Brower:1997ha,Singh:2019jog,Singh:2019uwd,Wiese:2014rla,Brower:1997ha,Brower:2020huh,Mathis:2020fuo,Halimeh:2020xfd, budde2024quantum, osborne2024quantum, Osborne:2023rzx,Luo:2019vmi}.

In this work, we consider gates that arise naturally when digitizing with the discrete subgroup approximation~\cite{Zohar:2014qma,Zohar:2016iic,Bender:2018rdp,Hackett:2018cel,Alexandru:2019nsa,Yamamoto:2020eqi,Ji:2020kjk,Haase:2020kaj,Carena:2021ltu,Armon:2021uqr,Gonzalez-Cuadra:2022hxt,Charles:2023zbl,Irmejs:2022gwv,Gustafson:2020yfe, Bender:2018rdp,Hackett:2018cel,Alexandru:2019nsa,Yamamoto:2020eqi,Ji:2020kjk,Haase:2020kaj,Carena:2021ltu,Armon:2021uqr,Gonzalez-Cuadra:2022hxt,Charles:2023zbl,Irmejs:2022gwv, Hartung:2022hoz,Carena:2024dzu, davoudi2024scattering,Lamm:2024jnl}. The discrete group approximation has advantages over the methods discussed above. It is a finite mapping of group elements to integers that preserves a group structure; therefore reducing the amount of fixed- or floating-point quantum arithmetic, simplifying the primitive gates. This method has its roots in early Euclidean LGT where the discrete group structure allowed for reduction of the classical resources~\cite{Creutz:1979zg,Creutz:1982dn,Bhanot:1981xp,Petcher:1980cq,Bhanot:1981pj,Weingarten:1980hx,Weingarten:1981jy} and has seen a resurgence in the era of quantum computation~\cite{Hackett:2018cel,Alexandru:2019nsa,Ji:2020kjk,Ji:2022qvr,Alexandru:2021jpm,Carena:2022hpz}. While the discrete subgroups are an approximation, they are related to the continuous groups broken by a Higgs mechanism~\cite{Kogut:1980qb,romers2007discrete,Fradkin:1978dv,Harlow:2018tng,Horn:1979fy} with work on-going to understand this systematically~\cite{Assi:2024pdn}.

 At present, most work has emphasized qubit devices, but recent demonstrations of multi-qudit gates, have increased interest in qudit-based digitization methods~\cite{Gustafson:2021jtq,Gustafson:2021qbt,Popov:2023xft,Calajo:2024qrc,Gonzalez-Cuadra:2022hxt, Illa:2024kmf, Zache:2023cfj, Murairi:2024xpc}. Given the abundance of platforms and native gates, it would be valuable to explore the relative merits of different gate sets with respect to algorithmic implementation. Qudit-based SRF architectures are one such platform, and we explore prototype gates required for the simulation of LGT within the discrete subgroup approximation on them here.  Focus is given to implementations on $d=4,6,8$ states -- respectively referred to as \emph{ququart}, \emph{quhexit}, and \emph{quoctit}. We consider both state preparation and gates decomposed in three ways: SNAP \& Displacement (S+D) gates, the echoed conditional displacement (ECD) \& single-qubit rotation gates $R(\theta,\phi)$, and optimal pulse control. Comparisons between the three are made in terms of number of blocks, time-scales, and noise resilience.

This paper is organized as follows. A brief review of the field theory motivations and the general protocols we are interested in are found in Sec.~\ref{sec:lattice}. In Sec.~\ref{sec:cqed}, we briefly summarize the theory behind coupling a cavity to a two-level system to produce a qudit. This is followed in Sec.~\ref{sec:gate-imp} with a discussion about how the interacting Hamiltonian is used to implement digital gates and the three decompositions.  Sec.~\ref{sec:methods} is where we describe the methods for determining our gate decompositions, the results of which are found in Sec.~\ref{sec:num-results} including an initial study of robustness to noise. We conclude in Sec.~\ref{sec:conc} with summary and future work.

\section{Lattice Gauge Theory}
\label{sec:lattice}
The gauge symmetries of particle physics constraint the possible interactions of theories, and thus only a finite set of gauge-group dependent primitive gates are required for simulation~\cite{Zohar:2016iic,Lamm:2019bik}. These correspond to operations that can be performed on either one or two elements $g,h$ of the group $G$ stored in registers $\ket{g},\ket{h}$.  For the case of pure gauge theory (theories without the  matter), the set of primitive gates are: the inverse gate $\mathfrak{U}_{-1}$ which sends a group element to its inverse:
\begin{equation}
\label{eq:inv}
    \mathfrak{U}_{-1}\ket{g} = \ket{g^{-1}};
\end{equation}
 the multiplication gate $\mathfrak{U}_{\times}$ that acts on two registers, changing  the target to the left product with the control:
\begin{equation}
\label{eq:mult}
    \mathfrak{U}_{\times}\ket{g}\ket{h} = \ket{g}\ket{gh};
\end{equation}
 the trace gate $\mathfrak{U}_{\rm Tr}$ which rotates $\ket{g}$ by a phase dependent on the trace of $g$ in a specified representation:
\begin{equation}
\label{eq:tr}
    \mathfrak{U}_{\rm Tr}\ket{g} = e^{i\theta Re\, Tr(g)}\ket{g};
\end{equation}
and the group Fourier transform $\mathfrak{U}_{FT}$ which acts on a single register with some amplitudes $f(g)$:
\begin{equation}
\label{eq:uft}
\mathfrak U_F \sum_{g \in G} f(g)\left|g \right>
=
\sum_{\rho \in \hat G} \hat f(\rho)_{ij} \left|\rho,i,j\right>.
\end{equation}
The second sum is over $\rho$, the irreducible representations of $G$; $\hat f$ denotes the $G$ Fourier transform of $f$,
\begin{eqnarray}
\hat{f}(\rho) = \sqrt{\frac{d_{\rho}}{|G|}} \sum_{g \in G} f(g) \rho(g),
\label{eqn:Fourier-group}
\end{eqnarray}
where $\vert G \vert$ is the group size, $d_{\rho}$ is the dimensionality of the representation $\rho$, and $f$ is a function over $G$.

Because these operations act on at most two registers, they can be analyzed and optimized more efficiently than working at the algorithmic level; something dramatically demonstrated for $\mathfrak{U}_{FT}$ in a recent work~\cite{Murairi:2024xpc}. Further, while $|G|$ (and therefore the optimal qudit dimension $d$) in the discrete subgroup approximation range from 24 to 1080, groups share structures which allows us to explore smaller gates while still being informative about preferred gate sets for HEP.

\begin{figure}
 \centering
        \includegraphics[width=0.32\linewidth]{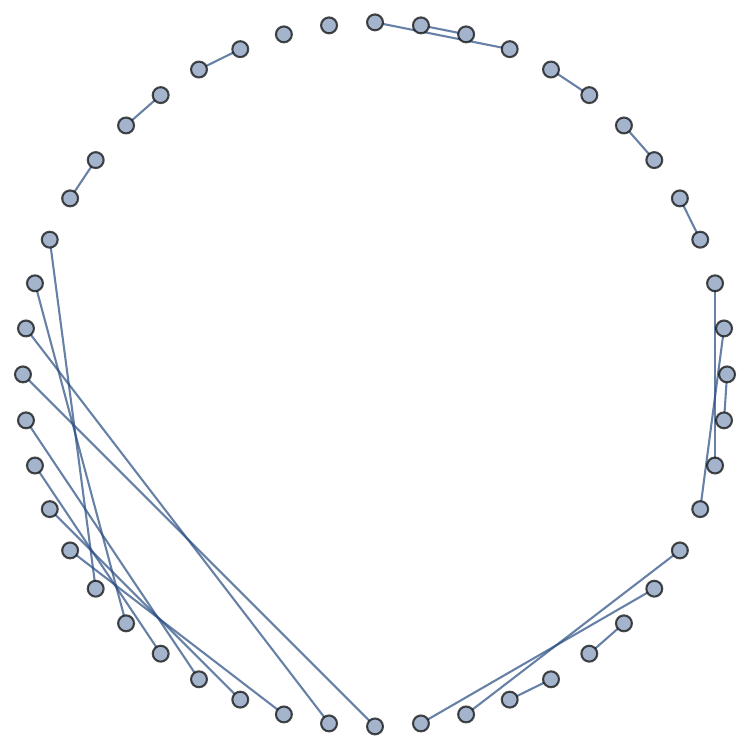}
        \includegraphics[width=0.32\linewidth]{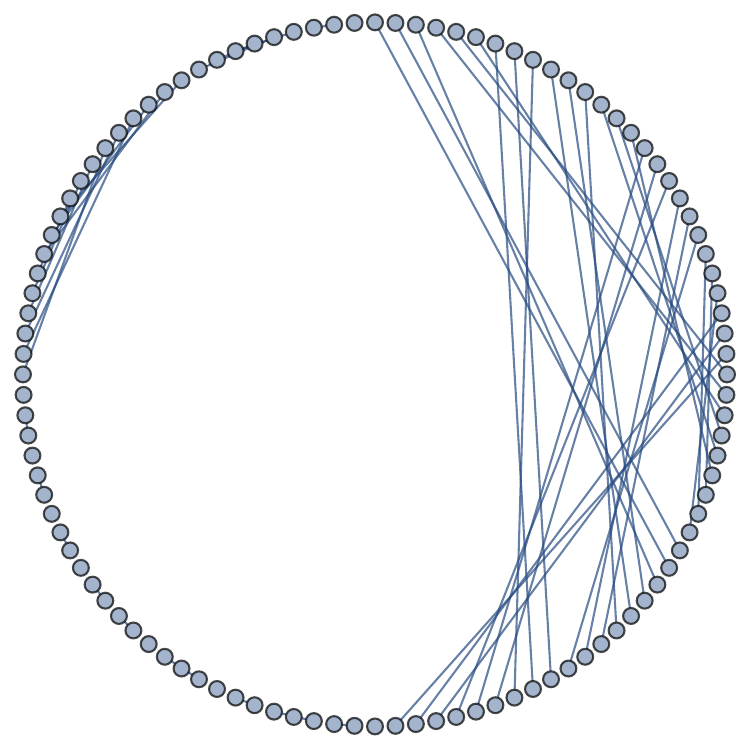}
        \includegraphics[width=0.32\linewidth]{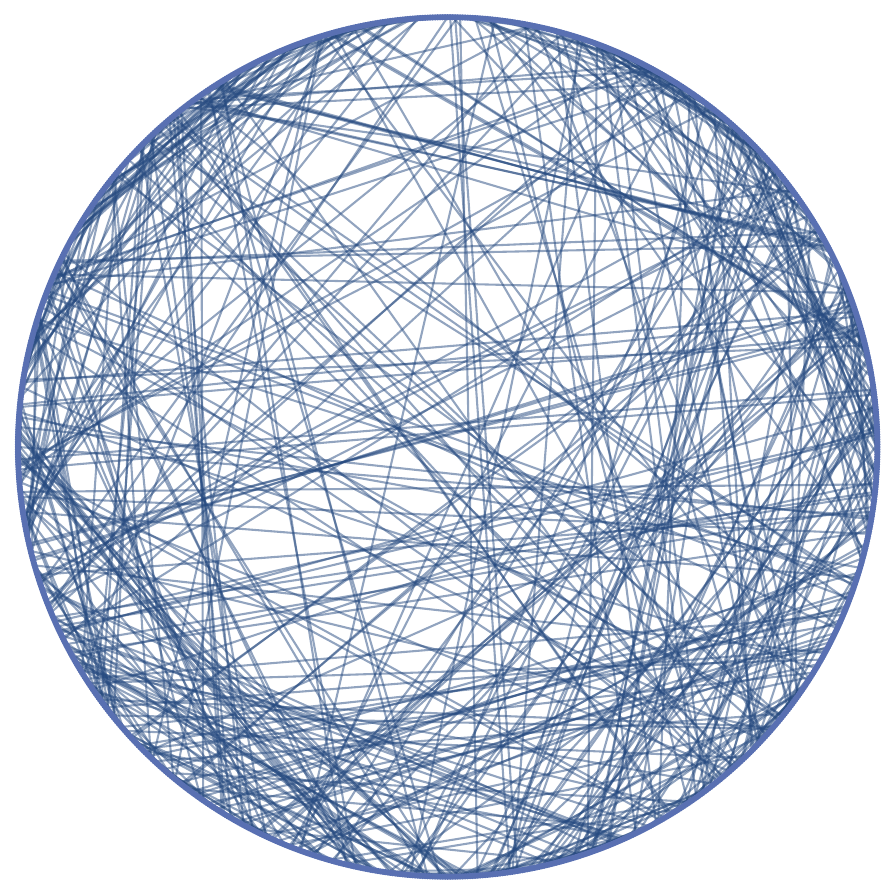}
        \caption{Permutation graph necessary for implementing $\mathfrak{U}_{-1}$ for different discrete group: (left) the 48 element $\mathbb{BO}$, (center) the 108 element $\Sigma{108}$, (right) the 1080 element $\Sigma(1080)$.}
    \label{fig:connect}
\end{figure}

\begin{figure*}
    \centering
        \includegraphics[width=\linewidth]{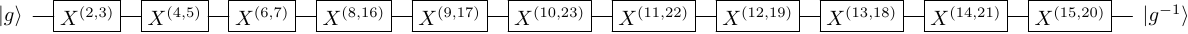}
        \vspace{0.1cm}
        
        \includegraphics[width=\linewidth]{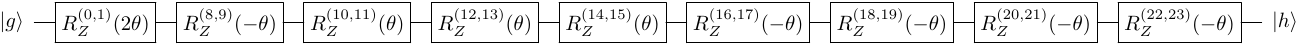}
        \caption{Implementation of $\btt$ primitive gate assuming a quicosotetrit: (top) $\mathfrak{U}_{-1}$  (bottom) $\mathfrak{U}_{\rm Tr}$. Note for the inverse gate the largest distance between a $\ket{g}$ and $\ket{g^{-1}}$ is 13}
        \label{fig:inverse24}
\end{figure*}

By inspecting Eq.~(\ref{eq:inv}), we recognize that $\mathfrak{U}_{-1}$ corresponds to a pair-wise permutation gate $\ket{g}\leftrightarrow\ket{g^{-1}}$ (See Fig.~\ref{fig:connect} for three example groups). If the qudit $d=|G|$, this gate is formed by a tensor product of at most $|G|/2$ Pauli $X^{(g,h)}$ gates between states $\ket{g},\ket{h}$.
% \begin{equation}
% \label{eq:x-gate}
%     X^{(g,h)}\left(a\ket{g} + b\ket{h}\right) = b\ket{g} + a\ket{h}.
% \end{equation}
An example of this for the 24 element $\mathbb{BT}$ mapped to a quicosotetrit ($d=24$ qudit) is shown in the top of Fig.~\ref{fig:inverse24}. As such, universal gate sets that can perform multiple $X^{(g,h)}$ concurrently would yield algorithmic improvement. If $d<|G|$, the gate structure generalizes from $X^{(g,h)}$ gates to requiring controlled permutations that couple the qudits.  Another important structure exemplified in Fig.~(\ref{fig:connect}) is that the maximum distance between $\ket{g}$ and $\ket{g^{-1}}$ is $\sim\frac{1}{2}|G|$ -- another way to discriminate the efficiency of gate sets.

The only two-register gate, $\mathfrak{U}_{\times}$, corresponds to a controlled permutation gate where a different permutation gate is applied depending on the control qudit state. As such, it has similar implementation requirements to $\mathfrak{U}_{-1}$. Further, $\mathfrak{U}_{\rm Tr}$ is diagonal in state space.  Therefore if $d=|G|$, the only elementary gate required is $R_Z^{(g,h)}(\theta)$ which generates a $R_Z$ gate of phase $\theta$ between $\ket{g}$ and  $\ket{h}$ (See the bottom of Fig.~\ref{fig:inverse24}).
% :
% \begin{equation*}
%     R_{Z}^{(g,h)}(\theta)(a\ket{g} + b\ket{h}) = ae^{-i\theta}\ket{g} + be^{+i\theta}\ket{h}.
% \end{equation*}

The final gate, $\mathfrak{U}_{\rm F}$, can be decomposed for $d=|G|$ into a tensor product of $SU(|\mathcal{C}|)$ rotations $\mathcal{U}^{(a,b\hdots, n)}_{N}$ where the size of set $\{\mathcal{C}\}$ is the order of the conjugacy class $\mathcal{C}$ of the group~\cite{Gustafson:2022xdt}.  For $d<|G|$, these must be mapped to controlled $SU(N)$ gates.  Although the spacing between states $a,b\hdots,n$ could in principle be $\sim|G|$, we find that for the discrete subgroup approximation, $|\mathcal{C}|\lesssim \sqrt{|G|}$ with the maximum $|\mathcal{C}|=18$. The smallest possible rotation is $SU(2)$, where the  Euler angle decomposition can be built from an ZXZ rotation where the superscripts indicate levels that are rotated:
\begin{equation}
    \mathcal{U}^{(a,b)}_2(\vec{\theta}) = R^{(a,b)}_Z(\theta_0) R^{(a,b)}_X(\theta_1)R^{(a,b)}_Z(\theta_2).
\end{equation}
We can combine $\mathcal{U}^{(a,b\hdots, n)}_{N}$ for the conjugacy class $\mathcal{C}$ into an operator $V_\mathcal{C}$.  For example, one conjugacy class with order 2 in $\mathbb{BT}$ yields~\cite{Gustafson:2022xdt}:
\begin{equation}
    \label{eq:vgm1}
    V_{1} = \prod_{a=0}^{11} \mathcal{U}_{2}^{(2a, 2a + 1)}(\vec{\theta}).
\end{equation}
$V_{\mathcal{C}}$ for larger $n$ are constructed recursively with additional Givens rotations~\cite{2002JPhA...3510467T, 2006JMP....47d3510B, 2004JGP....52..263T, 2005quant.ph.11019S}. $SU(3)$ rotations requires 
\begin{equation}
\begin{split}
    \mathcal{U}^{(a,b,c)}_3(\vec{\theta}) = &\, \mathcal{U}^{(a,b)}_2(\vec{\theta_0})
    R^{(b,c)}_X(\theta_1)\mathcal{U}^{(a,b)}_2(\vec{\theta_2})
     R_Z^{(b,c)}(\theta_3).
     \end{split}
\end{equation} 
while an $SU(6)$ needs from two $\mathcal{U}^{(a,b,c)}_3(\vec{\theta})$ and three $\mathcal{U}^{(a,b)}_2(\vec{\phi})$ (where we suppress the arguments below):
\begin{equation}
    \mathcal{U}_6^{(a,b,c,d,e,f)}=\mathcal{U}_3^{(a,b,c)}\mathcal{U}_3^{(d,e,f)}\mathcal{U}_2^{(a,d)}\mathcal{U}_2^{(b,e)}\mathcal{U}_2^{(c,f)}.
\end{equation}
With these, a $V_{\mathcal{C}}$ with order $6$ for $\mathbb{BT}$ is
\begin{equation}
\begin{split} 
    V_{9} = & \mathcal{U}_{6}^{(0, 9, 16, 1, 8, 17)} \mathcal{U}_{6}^{(2, 14, 20, 3, 15, 21)}\\
    &\mathcal{U}_{6}^{(4, 11, 23, 5, 10, 22)} \mathcal{U}_{6}^{(6, 12, 19, 7, 13, 18)}.
    \end{split}
\end{equation}
 Thus, $\mathfrak{U}_{\rm F}$ is a tensor product of a few $SU(N<18)$ rotations between well-separated qudit states.

Along with primitive gates, the opportunity for optimizing state preparation in LGT should be considered. In situations where gauge redundant encodings are necessary, one often prepares a gauge-dependent state and then applies a gauge-symmetrization operator. This gate acts as a projection operator~\cite{Zohar:2016iic,Stryker:2018efp,Lamm:2019bik,Carena:2024dzu}:
\begin{align}\label{eq:projection}
P \left|U_{ij}\hdots\right> &= \frac{1}{|G|^N}\left(\int_G \d g_1 \int_G \d g_2 \hdots\right)
\left|g_2 U_{ij} g^\dagger_1\hdots\right>\nonumber\\
&=
\frac{1}{|G|^N}\int_{G^N} \d g \; \phi(g) \left|U\right>.
\end{align}  The structure of this operator is a sum over $\mathfrak{U}_{\times}$, and thus again a set of controlled permutation matrices.  This constraint allows one to restrict the possible state that must be initially prepared, simplifying state preparation optimization.  Further, algorithmic choices in state preparation can simplify the possible initial states~\cite{Harmalkar:2020mpd,Gustafson:2020yfe,Saroni:2023uob}.

To summarize, the general structures of a LGT primitive gates are tensor products of $X^{(a,b)}$ and $SU(N<|G|)$ rotations between states that are typically well-separated. While it could prove experimentally hard to create qudits with $d=|G|$, for smaller $d$ the single register gates become similar to the controlled-permutations required for $\mathfrak{U}_{\times}$, so one can start with considering how prototype gates on smaller qudits can be compiled.

\section{3d cavity qed}
\label{sec:cqed}
The hardware platform we will study here corresponds to an SRF cavity mode (qumode) coupled to a two level system (qubit).  With $\hbar=1$, a simplistic model of this interaction from circuit QED is the \emph {Jaynes-Cummings model}~\cite{Blais:2020wjs}:
\begin{equation}
\label{jaynes-cummings}
    H =  \omega_c \hat{a}^\dagger\hat{a} + \frac{1}{2} \omega_0 \sigma_z +  g  (\hat{a}^\dagger\sigma_- + \hat{a}\sigma_+).
\end{equation}
where $\omega_c$ is the frequency of the qumode with creation and annihilation operators $\hat{a},\hat{a}^\dagger$, $\omega_0$ is the frequency of the two-level system which is acted on by the Pauli operators $\sigma_i$, and $g$ is a coupling between the qumode and qubit.
The regime considered, particularly amenable to hardware, is the so-called dispersive regime, which occurs when $\Delta = |\omega_0 - \omega_c| \gg g$~\cite{dispersive-regime}. In this regime, one can approximately diagonalize $H$ using a unitary transformation $D^\dag HD$ with 
\begin{equation}
    D=e^{\lambda(\hat{a}^\dagger\sigma_- - \hat{a}\sigma_+)}
\end{equation} where $\lambda=g/\Delta$. By doing so, and expanding to $\mathcal{O}(\lambda)$, one obtains
\begin{equation}
\label{jaynes-cummings-final}
    H =  \omega_c \hat{a}^\dagger\hat{a} + \frac{1}{2}\omega_q \sigma_z + \frac{1} {2}\chi\hat{a}^\dagger\hat{a}\sigma_z+\mathcal{O}(\lambda^2) , 
\end{equation}
where $\chi = 2g\lambda=2g^2/\Delta$ is called the dispersive shift and $\omega_q = \omega_0 + \chi$. With this, the Fock states of the qumode ($\ket{0}, \ket{1}, \ket{2}$ ...) can be used as a qudit computational space, and the ancilla qubit provides an interaction to manipulate the qudit. 

An easy way to control the system is to add additional fields acting on the qumode and the qubit. A possible choice is to add a single control on both:
\begin{align}
\label{eq:jc-with-pulses}
    H =  &\,\omega_c \hat{a}^\dagger\hat{a} + \frac{1}{2}  \omega_q \sigma_z +  \chi\hat{a}^\dagger\hat{a}\sigma_z \notag\\&+  \epsilon(t) \hat{a}^\dagger +  \epsilon(t)^*\hat{a} +  \Omega(t)\sigma_+ +  \Omega(t)^*\sigma_-.
\end{align}
which we define as $H = H_0 + H_1$ as a non-interacting $    H_0 = \omega_c \hat{a}^\dagger\hat{a} + \frac{1}{2} \omega_q \sigma_z$, and the remaining terms the interacting Hamiltonian $H_1$. 

With this choice of controls and enough time, it is possible to manipulate the system to reach any arbitrary state~\cite{PhysRevA.92.040303}.
It is important to emphasise that for SRF cavities there is not an agreed-upon mode of operation. Therefore, the investigation of different gate sets \cite{krastanov_snap, krastanov2, Kudra_2022, Eickbusch:2021uod} and optimal control \cite{PhysRevA.92.040303, Strauch_2012, PhysRevLett.76.1055, roque2020engineering, Fei:2022njl, Seifert:2022nxm, luchi2023control,Reineri:2023ebw,You:2024oru}, is key to fully exploit these systems. At a higher level of abstraction, performing efficient compilation of quantum circuits to qudits rather the qubits also requires new techniques~\cite{Job:2023huu,Ogunkoya:2023ryv,Roy:2023alp,Mato:2023odr,Ogunkoya:2023cfg,Gustafson:2024kym}.

\section{Gate implementation}
\label{sec:gate-imp}
 To obtain a theoretical target gate $U_t$, one manipulates the time-dependent couplings of Eq.~(\ref{eq:jc-with-pulses}). All approaches require this optimization at the Hamiltonian level, and in this section we summarize two ways to do so. The first approach only relies on direct optimization of $\epsilon(t),\Omega(t)$, while the other decomposes $U_t$ into a set of simpler gates that have been optimized at the Hamiltonian level. In either case, one considers a quantum state $\ket{\psi(t)}$ evolving according to the Schr\"odinger equation with a particular time-dependent Hamiltonian $H(t)$. This state can be approximated by dividing $t$ into time steps of size $\delta t$: 
\begin{equation}
    \label{eq:trotter}
    \ket{\psi(t)} \approx \prod_{n=0}^{t/\delta t }e^{-i H(n \delta t)\delta t } \ket{\psi(0)} \equiv V\ket{\psi(0)},
\end{equation}
with increasing accuracy as $\delta t\rightarrow 0$.

Clearly, $H(t)$ allows transition between quantum states, and thus $V$ correspond to a digital quantum circuit.  The goal is then is to find a $V$ which implements $U_t$ with high fidelity. It is important to emphasise that $V$ as defined in Eq.~(\ref{eq:trotter}) makes reference to a state $\ket{\psi(0)}$. It is the case that the state-dependent fidelities vary substantially~\cite{Sahinoglu2020hamiltonian,Hatomura:2022yga,Burgarth:2022lib} and therefore synthesizing a general unitary is harder than preparing a state from a given $\ket{\psi(0)}$.
Proceeding, we will distinguish between gates for \emph{state preparation} which are optimized only for acting on vacuum $\ket{0}$, and general gates which should have high fidelity on any $\ket{\psi(0)}$.

 To study these effects, we need to define the infidelity i.e. a cost function $\mathcal{I}(U_t, V)$ to minimize. Given that the goal is to implement a target gate $U_t$,  a common approach consists of dividing the time into $N$ time steps and for each of these, the control pulses $\Omega(n\delta t), \epsilon(n \delta t)$ are constant. A main drawback of this approach is the increasing number of time steps -- and therefore optimization parameters-- required when increasing $d$ of the qudit~\cite{PhysRevA.92.040303,Eickbusch:2021uod}, this is because while the controls can generate the algebra, reaching higher states requires more and more commutators. Another drawback is the pulse discontinuities. By optimizing only the field amplitudes at each time step, we are neglecting the time required to ramp the fields on the physical hardware, which can introduce additional infidelity or must be implemented as expensive constraints on the optimization. Further, increasing $N$ requires more matrix multiplications in the optimization algorithm. Looking at \eref{eq:trotter}, for a $d$-dimensional qudit $N$ matrix multiplications are required to compute the evolution of the system, leading to $\mathcal{O}(Nd^3)$ operations. Lastly, when optimizing the cost function, unless the optimization method is gradient-free, it is necessary to compute either the gradient or in some cases the hessian -- further increases the complexity. For $N$ parameters, without recycling computations, the cost function needs to be evaluated $N+1$ times, which leads to an overall complexity of $\mathcal{O}(N^2d^3)$. 

An alternative approach avoids piecewise constant terms and instead optimize gates in a different basis. Looking at the expression of \eref{eq:jc-with-pulses}, a way to generate smooth pulses, making it easier to implement, is to expand them in some polynomial basis~\cite{kormann:2010aaa,Pagano:2024rwb}, where the optimal choice can be problem-dependent. Moreover, the number of basis states directly controls how steep the pulse ramping can be. Here, we will expand in Chebyshev polynomials which provide an exact $n^{th}$ order interpolation function for any $n^{th}$ order polynomial. Approximating instead using the Fourier series requires more terms to reach a comparable accuracy \cite{karjanto2020properties}. Rewriting \eref{eq:jc-with-pulses} in the Chebyshev basis of $T_k(t)$, yields

\begin{align}
\label{eq:chebyshev-hamiltonian}
    H = H_0 + \chi\hat{a}^\dagger\hat{a}\sigma_z +  \sum_{k=0}^{N}\left[ (c_{k} \hat{a}^\dagger +q_{k}\sigma_+)T_k(t)  + h.c.\right],
\end{align}
where $\{c_k\}$ and  $\{q_k\}$ are the coefficients of the qumode and the qubit, respectively. Using this expression for $H$, it is possible to compute the $V$. Its magnitude depends on the dimension of the cavity space that is optimized, the duration of the pulses, and on the resolution of the hardware on which the pulses are optimized.

\subsection{Interaction picture}
Finding an optimal $V$ from \eref{eq:chebyshev-hamiltonian} can be made easier by transforming into the interaction frame of reference~\cite{gerry_knight_2004}. Since $H_0$ is time-independent, the time-evolution can be solved analytically: 
\begin{equation*}
    U_0(t) = e^{-iH_0t}.
\end{equation*}
With this expression, it is possible to define a state vector in this new frame of reference:
\begin{equation}
\label{eq:int-transformation}
    \ket{\psi_I(t)} = U_0(t)\ket{\psi_S(t)},
\end{equation}
where $\ket{\psi_S(t)}$ represents the state governed by Schrodinger equation.
Using this, it is possible to write

\begin{align}
\label{eq:sch}
     i \frac{\partial}{\partial t}&\ket{\psi_S(t)} = H \ket{\psi_S(t)} = e^{-iH_0t} \left(H_0 + i\frac{\partial}{\partial t}\right) \ket{\psi_I(t)}\notag \\ 
     &\implies i \frac{\partial}{\partial t}\ket{\psi_I(t)} = e^{iH_0t} H_1 e^{-iH_0t}\ket{\psi_I(t)}.
\end{align}

This can be used to simplified optimization of $V$ by solving Eq.~(\ref{eq:sch}) and then transforming back to the initial frame by multiplying by the inverse transformation defined by \eref{eq:int-transformation}.
The remaining step consists in writing explicitly the term that dictates the evolution of the system: $$H_{1_{I}} = e^{iH_0t} H_1 e^{-iH_0t}$$
with the resulting operator being
\begin{equation}
\label{eq:int-hamiltonian}
H_{1_{I}} = \chi\hat{a}^\dagger\hat{a}\sigma_z +  \sum_{k=0}^{N} [c_{k} C_k(t) \hat{a}^\dagger+q_{k} Q_k(t) \sigma_+  + h.c].
\end{equation}
where 
\begin{equation*}
  \begin{split}
C_k(t)\equiv T_k(t) e^{i\omega_c t} \\ 
 Q_k(t)\equiv T_k(t) e^{i\omega_q t},
\end{split}  
\end{equation*}

While optimizing the pulses in the interaction picture may be easier, the pulses must be converted back to the laboratory frame by multiplying them with the exponential factors that are missing in \eref{eq:int-hamiltonian}. This may reduce accuracy, depending on the hardware precision.

To inform the gate synthesis that is developed in this work, it is important to give some realistic near-term properties for 3D cavities. We take for a fiducial decoherence time of the cavity $T_1 =10~ms$~\cite{Alam:2022crs}. With this, we take $\omega_c=5$ GHz, $\omega_q=3$ GHz, and $\chi=0.3$ MHz. Regarding the control pulses, we consider a max time resolution of 1 ns (corresponding to 1 giga-sample/$s$) for the state preparation, and a longer 10ns time resolution for the pulses used for gate optimization. An additional constraint is for the drive amplitudes to change by no more than a 1 MHz/$ns$ (to maintain a smooth drive profile) which is consistent with near-term SRF cavities.

\subsection{Gate decomposition in universal sets}
One could imagine engineering an optimized pulse for each LGT primitive, although issues with computational expense and error correction may complicate this.  Instead, one might decompose each primitive into a limited set of universal gates whose implementation is known at the Hamiltonian level with high fidelity. The price of this is increased runtime of the primitive. While qudits can decrease the circuit depth for LGT~\cite{Gustafson:2022xdt}, improvements are gate-set dependent.  Thus we consider two universal sets: SNAP \& Displacement and ECD.

\subsubsection{SNAP \& Displacement}

The qudit gate set of SNAP and Displacement gates ($\sdg$) has been widely discussed in the literature \cite{krastanov_snap, krastanov2, Kudra_2022,Job:2023huu} and we briefly review the key points here.
Crucially, the universality of $\sdg$ is known~\cite{PhysRevA.92.040303}.
The Displacement gate is produced by driving the cavity on resonance with a calibrated pulse of a specific amplitude and length.  It can be expressed as
\begin{equation}
\label{disp}
    D(\alpha) = e^{\left(\alpha \hat{a}^\dagger - \alpha^* \hat{a}\right)},
\end{equation}
and generates a coherent state $\ket{\alpha}$ from the vacuum displaced in phase space by $\alpha$ i.e. $D(\alpha)\ket{0} = \ket{\alpha}$.  An important property of $D(\alpha)$ is that its Hermitian conjugate $D^\dag(\alpha)=D(-\alpha)$.  The parameter $\alpha$ of \eref{disp} can be considered real, without loss of generality.  Mathematically $D(\alpha)$ represents the Heisenberg group which may prove useful for compilation. For reasonable $\chi$ on GHz SRF cavities with $T_1,T_2\sim \mathcal{O}(1)$ ms, the gate times range from $0.01-0.2$ $\mu$s depending on $\alpha$~\cite{axline2018demand}.

The SNAP gate applies arbitrary phases to Fock states
\begin{equation}
\label{snap}
    S(\Vec{\theta}) = \sum_{k=0}^{d-1} e^{i\theta_k} \ket{k}\bra{k}.
\end{equation}
This is produced by driving the qubit with $\pi$ pulses with relative phases, which cause a phase kick-back onto the qudit. Importantly, while the qubit is used to implement this gate, upon completion the qubit is unentangled from the qudit.  To ensure high fidelity, the driving strength for a SNAP must be taken $\max(Q_k(t)) \ll \chi$, with a good heuristic found to be a pulse length of $\pi/(10\chi)$. While larger $\chi$ allows for faster SNAP gates, the need to avoid strong self-Kerr interactions limits its size. Reasonable values for $\chi$ for GHz cavities with $T_1,T_2\sim 1$ ms lead to SNAP gate times between $1-50$ $\mu$s~\cite{Alam:2022crs,Eickbusch:2021uod,Roy:2024uro}. 

With these, any single-qudit gate can be approximately decomposed by~\cite{krastanov2}:
\begin{equation}
\label{eq:snap1}
    U_t = \left(\prod_{k=0}^M D(\alpha_k)S(\Vec{\theta}_k) \right) D(\alpha_{M+1}),
\end{equation}
where $B$ is the number of blocks used. For a $d$-state qudit, an $B$ block decomposition has $(d+1)B+1$ free parameters.  Since an arbitrary qudit state can be obtained by an $SU(d)$ rotation with $d^2-1$ parameters, a naive estimate for the decomposition of any single qudit state is at most $B=\frac{d^2-2}{d+1}\sim \mathcal{O}(d)$~\cite{Fosel:2020oyj}. This represented an algorithmic improvement over the $\mathcal{O}(d^2)$ scaling of a $\log_2(d)$ qubit device with access to only one and two-qubit gates~\cite{Mansky:2022bai}. The authors of~\cite{krastanov2} empirically found that a broad set of qudit gates (with $d\leq 10$) can be decomposed according to Eq.~(\ref{eq:snap1}) with only 3 to 4 blocks and that the fidelity $\mathcal{F}\approx 1-e^{-6.49(B/d)^{1.91}}$.

\begin{figure}
    \centering
    \includegraphics[width=\linewidth]{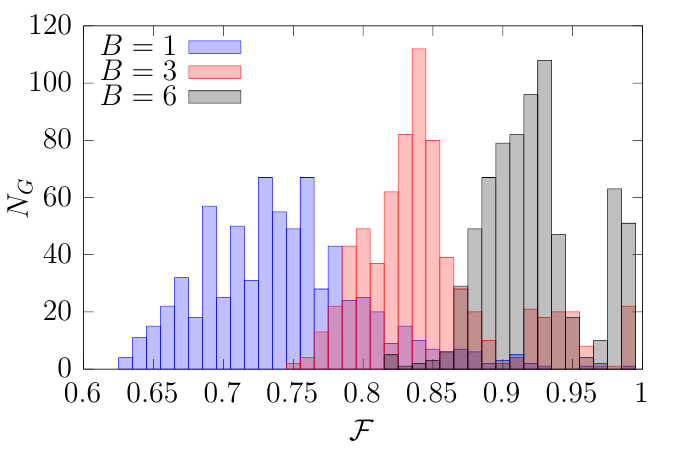}
        \caption{Number of $d=6$ permutation gates $N_G$ with given fidelity as a function of $B$ for $S+D$.}
    \label{fig:histo_perms}
\end{figure}

\begin{figure}
        \includegraphics[width=\linewidth]{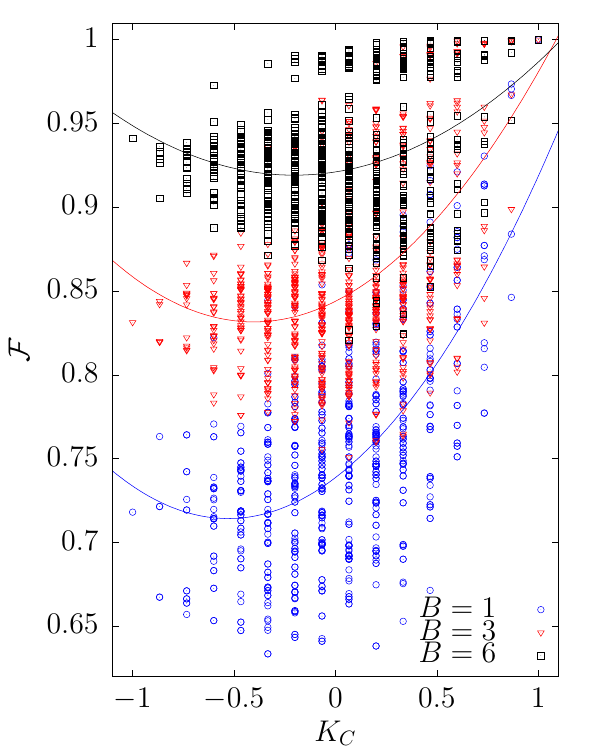}
    \caption{Fidelity of $d=6$ permutation gates vs. $K_C$ and $B$. Quadratic best fit lines are shown to guide the eye.}
    \label{fig:ktau_perms}
\end{figure}

Following the LGT insight discussed in Sec.~\ref{sec:lattice} that permutation matrices represent a sizeable fraction of the primitives necessary, we investigated whether these gates represent a special subset of all qudit gates in terms of their fidelity under $S+D$ decomposition.  For all d! permutation gates with d = [3, 6] we performed 10 decompositions with $B$ = [1, 6], where the parameters for the decompositions were randomly initialized.  From this data, the best fit to the functional form advocated in ~\cite{krastanov2} was $\mathcal{F_{\rm perm}}\approx 1-e^{-3.8(5)(B/d)^{0.5(1)}}$.  This dependence shows that for values of $B/d\lesssim 1$, permutation gates can be decomposed with higher fidelity than general gates.  In Fig.~\ref{fig:histo_perms} we present the distribution of permutation gate fidelities for $d=6$ as a function of M.  From the figure, one observes a clear bimodal distribution in $\mathcal{F}$ for $B=3,6$.
Beyond finding that permutation gates are easier to decompose than general $SU(d)$ matrices, one could ask if specific permutation decompositions are better -- especially in light of the bimodal distribution.  Given the freedom in encoding of LGT, such algorithmic input could be used similar to Gray codes on qubits to reduce noise or gates~\cite{2004PhRvL..92q7902V,2020npjQI...6...49S,2021PhRvA.103d2405D,2021arXiv210308056C}.  As such, we investigated the dependence of $\mathcal{F}$ for $d=6$ permutation gates on two distance metrics used for distinguishing between strings when we write the permutations as tuples, for example
\begin{equation}
    \begin{pmatrix}
        0 & 1 & 0\\
        1 & 0 & 0\\
        0 & 0 & 1\\
    \end{pmatrix}\rightarrow (1,0,2).
\end{equation}
where the $i-$th entry $j$ of the tuple corresponds to permuting the state $\ket{j}\rightarrow\ket{i}$.
The first is the \emph{Kendall rank correlation coefficient} $K_C$ which is used as a statistic measure between two sets of data $x=\{x_0\hdots x_n\}$ and $y=\{y_0\hdots y_n\}$ of their relative monotonicity~\cite{10.1093/biomet/30.1-2.81}.  It is defined in terms of the number of pairs $(x_i,y_i)$ and $(x_j,y_j)$ that are concordant, i.e. for $i<j$ either $x_i<x_j$ and $y_i<y_j$ or $x_i>x_j$ and $y_i>y_j$.  If we call this number $n_c$, and the remaining discordant pairs $n_d$ then $K_C$ is 
\begin{equation}
K_C=\frac{n_c-n_d}{n_c+n_d}
\end{equation}
where $K_C$ can range from 1 when the two sets are identical to -1 when the two sets are exactly reverse. We take $x=(0,1\hdots, d)=\mathbb{1}_d$, and thus $K_C$ is measured from it. The results are found in Fig.~\ref{fig:ktau_perms}.  
From these it is clear that $K_C$ does capture some of the dependence of $\mathcal{F}$ on permutation, with higher $K_C$ typically having larger $\mathcal{F}$.  In particular we see that the bimodal distribution of Fig.~\ref{fig:histo_perms} is correlated with positive $K_C$.  Investigating further, one observes that $\mathcal{F}(K_C)$ is non-monotonic, reaching a minimum at $K_C=0$ with a large variance, before rising slightly toward $K_C\rightarrow-1$.  

\begin{figure}
        \includegraphics[width=\linewidth]{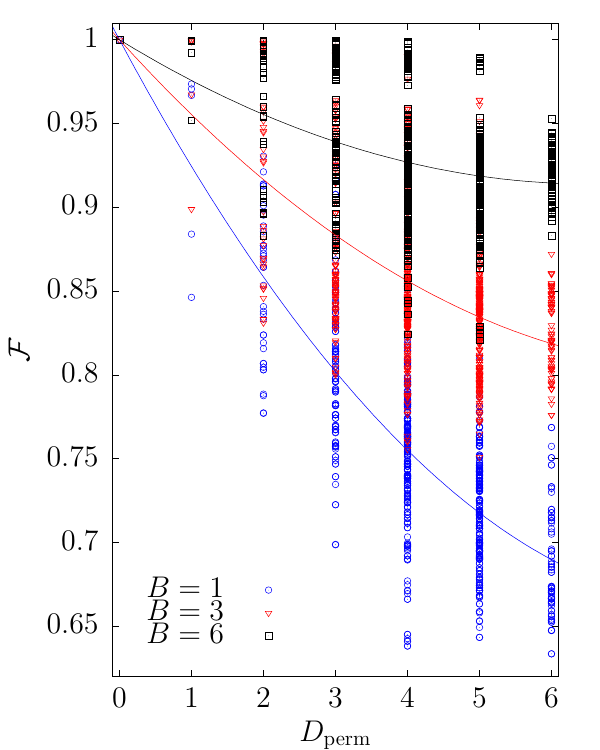}
    \caption{Fidelity of $d=6$ permutation gates vs. $D_{\rm perm}$ for different block sizes $B$. Quadratic best fit lines with $\mathcal{F}(0)=1$ are shown to guide the eye.}
    \label{fig:ddl_perms}
\end{figure}

Another metric for comparing sets, with its origin in spell checking, is the \emph{Damerau–Levenshtein distance}~\cite{10.1145/363958.363994,levenshtein1966binary} $D_{\rm perm}$ which here counts only the number of transposition of two adjacent characters required to change one word into another.  Fig.~\ref{fig:ddl_perms} presents $\mathcal{F}$ as a function of $D_{\rm perm}$. We do observe decreasing $\mathcal{F}$ with $D_{\rm perm}$. From the best fit lines in Fig~\ref{fig:ddl_perms}, we see that $1-\mathcal{F}\approx[10^{-3},10^{-2}]D_{\rm perm}$ with some weak quadratic term. With this, one can make a comparison between directly decomposing a gate with and decomposing in terms of $N=D_{perm}$ single $X^{(a,a+1)}$ with smaller $B$ each with individual $\mathcal{F}_{ind}$.  In the best case for the single gates,
\begin{equation}
    1-10^{-2}D_{\rm perm}=\mathcal{F}_{ind}^{D_{\rm perm}}\implies F_{ind}>0.990
\end{equation}
Since this high fidelity for $B<D_{perm}$ decomposing directly into $S+D$ is always more efficient.

Taken together, we conclude that permutation gates represent an important subset of gates under $S+D$ decomposition with higher than average fidelity.  Further, specific classes of permutation gates appear to have still better scaling and thus could be useful in optimizing digitization of LGT.

\subsubsection{ECD decomposition}
One limitation of S + D is that  operations require time comparable to $2\pi / \chi$. This is a limiting factor in the weak-dispersive regime because decoherence will reduce fidelity. To overcome this challenge, another universal set has been proposed which combines an echoed conditional displacement $ECD$ gate with single qubit rotations~\cite{Eickbusch:2021uod}:
\begin{equation}
\label{eq:ecd-decomposition}
    U_t = \left(\prod_{k=0}^M R(\theta_k, \phi_k) ECD(\alpha_k)\right)R(\theta_M, \phi_M),
\end{equation}
where the ECD gate corresponds to applying $D$ gates conditioned on the state of the qubit,
\begin{equation*}
    ECD(\alpha) = D(\alpha/2)\ket{e}\bra{g} + D(-\alpha/2)\ket{g}\bra{e}.
\end{equation*}
To enable universality, this is coupled to unselective qubit rotation 
\begin{equation*}
    R(\theta, \phi) = \exp\left(–\frac{i\theta}{2}(\sigma_x \cos \phi+ \sigma_y \sin \phi)\right).
\end{equation*}
In constrast to S+D, only the $ECD$ gate acts directly on the qudit.  In this scheme $R(\theta, \phi)$ acts on the qubit changing the relative displacement. As such, the qubit and qudit may still be entangled at the end of a block.  This leads to greater sensitivity of the qudit to noise from the qubit. Modern transmon qubits can perform $R(\theta, \phi)$ in $\mathcal{O}(0.01$ $\mu$s). $ECD(\alpha)$ can be implemented in $T\propto\frac{\alpha}{\chi\alpha_0}$ where $\alpha_0$ is the magnitude of the intermediate oscillator displacement. For GHz cavities and reasonable values of $\alpha_0\sim 30$ this would correspond to $0.1-0.5$ $\mu$s~\cite{Eickbusch:2021uod,Roy:2024uro}. 

Counting parameters, $B$ blocks have $3B+2$.  Since the qubit may not be decoupled, the general state is given by an $SU(2d)$ rotation with $4d^2-1$ parameters.  This naively estimates that $\frac{4}{3}d^2-1$ blocks are needed -- $\mathcal{O}(d^2)$ scaling  -- which is parametrically worse that $S+D$.
As will be seen below, this larger $B$ can be compensated by the faster gate times.

One should also compare to State-of-the-art qubit implementations, which also scale asymptotically as $\mathcal{O}(d^2)$. An $SU(2^n)$ unitary where  $d=2^n$ requires~\cite{Mansky:2022bai} a number of sequential CNOTs
\begin{equation}
    N_{\rm CNOT}=\frac{21}{16}d^2-d-\frac{3}{4}d\log_{2}(d)
\end{equation} 
which suggests a comparable circuit depth to $ECD$. Further, the theoretical lower bound is $N_{\rm CNOT}=\frac{1}{4}[d^2-1-3\log_2(d)]$~\cite{PhysRevA.69.062321} while the lower bound on $ECD$ is unknown. Importantly, modern transmon-based platforms can perform CNOT gates in $\sim 0.5~\mu s$~\cite{McKay:2023nxa}, similar to the higher estimates for near-term ECD gates. Taken together, these results suggest that the algorithmic advantage of qudit-based platforms with ECD gate sets over qubit devices depends upon acheiving gate times below $0.5~\mu s$ and clearer decomposition estimates.

\section{Methods}
\label{sec:methods}
For optimization, we have used \texttt{LBFGS} as our optimization algorithm \cite{liu1989limited}. For preparing $\ket{\Psi}$, one solves the pulse optimization of \eref{eq:trotter} using the parameters in \eref{eq:chebyshev-hamiltonian} with a fixed $\ket{\psi(0)}$. For general gate preparation, the more complicated optimization procedure motivates using the interaction picture of \eref{eq:int-hamiltonian}. Further, given the larger number of optimization parameters, we found efficiency gains from first optimizing pulses with a small number of coefficients, and then growing the bases while using as initial conditions for the next iterations the previous results\footnote{The entire parallelized codebase is written in \texttt{Julia} and can be found at \url{https://github.com/AndreaMaestri18/Opt3DQalgs}.}
In order to compare the optimized gates with the target, it is necessary to define a cost function. In the literature, there is a number considered, typically related to $p$ norms~\cite{Fei:2022njl, Gustafson:2022xdt, Eickbusch:2021uod, PhysRevLett.76.1055, PhysRevA.92.040303, krastanov_snap, krastanov2, luchi2023control}:
\begin{equation}
    ||A||_p=\sup_{x\neq 0}\frac{||Ax||_p}{||x||_p}
\end{equation}
which use the $p$ vector norm,
\begin{equation}
    ||x||_p= \left(\sum_i |x_i|^p\right)^{1/p}.
\end{equation}

If the goal is preparing $\ket{\Psi}$, the optimization problem can be restricted to only a subspace of the full system. Here, we follow prior work and use as a cost function the infidelity:
\begin{equation}
\label{eq:norm-state-prep}
\mathcal{I} = 1-|\bra{\Psi} V \ket{\psi(0)}|^2,
\end{equation}
Typically, $\ket{\psi(0)}=\ket{0}$, but we will investigate a limited set of other initial states as a way to understand robustness of state preparation to noise.

In case where we wish to optimize $V$ to approximate a gate $U_t$, one must account for working in the interacting picture by multiplying the target by the inverse of the transformation $U_0(T)$ defined in \eref{eq:int-transformation}.  With this, the most common cost function to minimize is :
\begin{equation}
\label{eq:trace-norm}
 \mathcal{I} = 1 - \left(\frac{|\tr(U_t^\dagger U_0(T)V)|}{2d}\right)^2
\end{equation}
where $2d$ is the dimension of the combined cavity and qubit space. The idea behind this measure is that if $V\approx U_t$, using the properties of the unitary matrices we have $\mathcal{I} \longrightarrow 0$. Other options to measure the distance between 2 matrices could be:
\begin{equation}
\label{eq:oneminusnorm}
 \mathcal{I} = ||\mathbb{1}- U_{t}^\dag U_0(T) V||_2.
\end{equation} which still relies on unitarity, and the norm of the difference between the matrices:
\begin{equation}
\label{eq:norm}
\mathcal{I} = ||U_{t} - U_0(T)V||_2.
\end{equation}

\section{Numerical Results}
\label{sec:num-results}
\begin{figure}
    \centering
    \includegraphics[width=\linewidth]{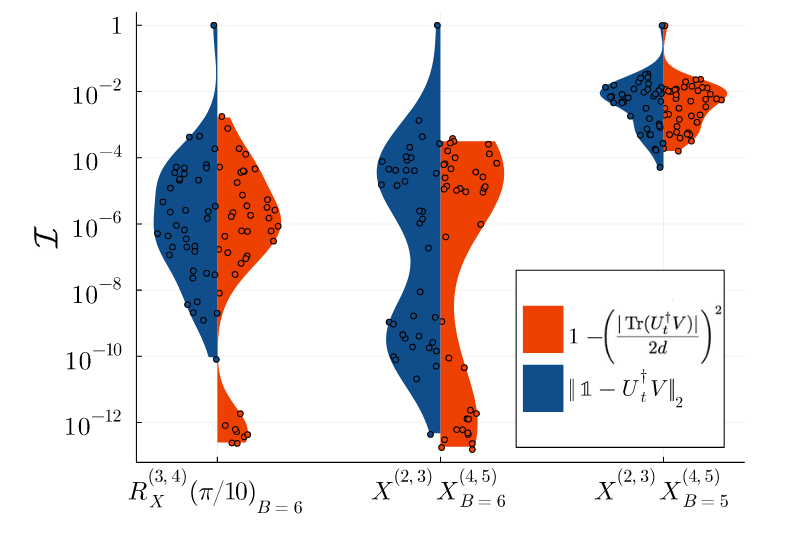}
    \caption{Distribution of $\mathcal{I}$ of two cost functions for three gates on a quhexit using $S+D$ decomposition, after 30 optimizations and using different numbers of blocks $B$.}
    \label{fig:cost-function-analysis}
\end{figure}

In this section, we numerically optimize decompositions of a set of prototypical gates for HEP.  As a first step, we studied how minimizing the different infidelities $\mathcal{I}$ of Eqs.~(\ref{eq:trace-norm}),~(\ref{eq:norm}), and~(\ref{eq:oneminusnorm}) as a cost function. This preliminary analysis was used to decide which cost function going forward. The cases considered were $S+D$ decomposition of two gates on a quhexit: a double $X$ gate, $X^{(2,3)}X^{(4,5)}$ (with two different block sizes) and $R_X^{(2,3)}(\pi / 10)$ gate. While we tested the three norms, the results coming from the optimization of \cref{eq:norm} showed poor, slow convergence and thus in-depth analysis was done only on the remaining cost functions which are presented in Fig~\ref{fig:cost-function-analysis}.  In this figure, we present the infidelity of trace-norm for both optimization functions~\cref{eq:oneminusnorm}.   

 One observes from these results that varying cost function leads to different decompositions, with distinct distributions related to how the matrix elements are weighted. A more sophisticated two-sided t-test for unequal variances confirms this. Putting everything together, the trace distance defined in \eref{eq:trace-norm} generally produced lower $\mathcal{I}$. Therefore this metric will be used in what follows. 
 
We now study preparation of the following states $\ket{\Psi}$ on all three qudits and using all three decompositions:
\begin{itemize}
    \item Fock state 3: $a^3\ket{0}=\ket{3}$, which is a representative of an electric basis state
    \item $d-$Hadamard state: $H_d\ket{0}=\frac{1}{\sqrt{d}}\sum_{i}\ket{i}$, which can be taken as a prototype of the weak-coupling gauge-invariant vacuum state in the magnetic basis
    \item $X^{(3,4)}$, $R_X^{(3,4)}(\pi/10)$, $U_2^{(3,4)}(\frac{\pi}{5}, \frac{2\pi}{15}, \frac{\pi}{10})$ applied to randomly-generated state: $\ket{\psi_{ran}} \rightarrow U_{target}\ket{\psi_{ran}}$, correspond to prototypes of general HEP states that are prepared on a potentially noisy device.
\end{itemize}

\begin{figure}
    \centering
\includegraphics[width=\linewidth]{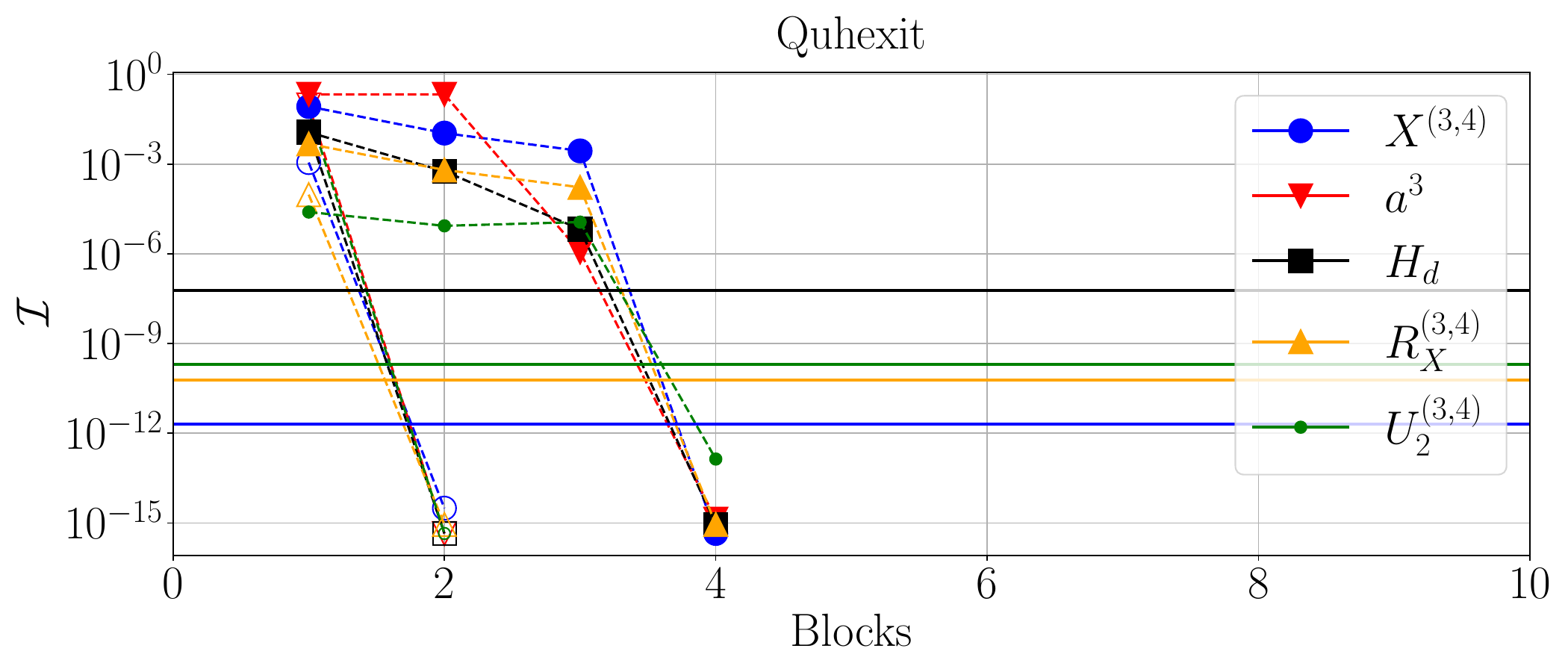}
\includegraphics[width=\linewidth]{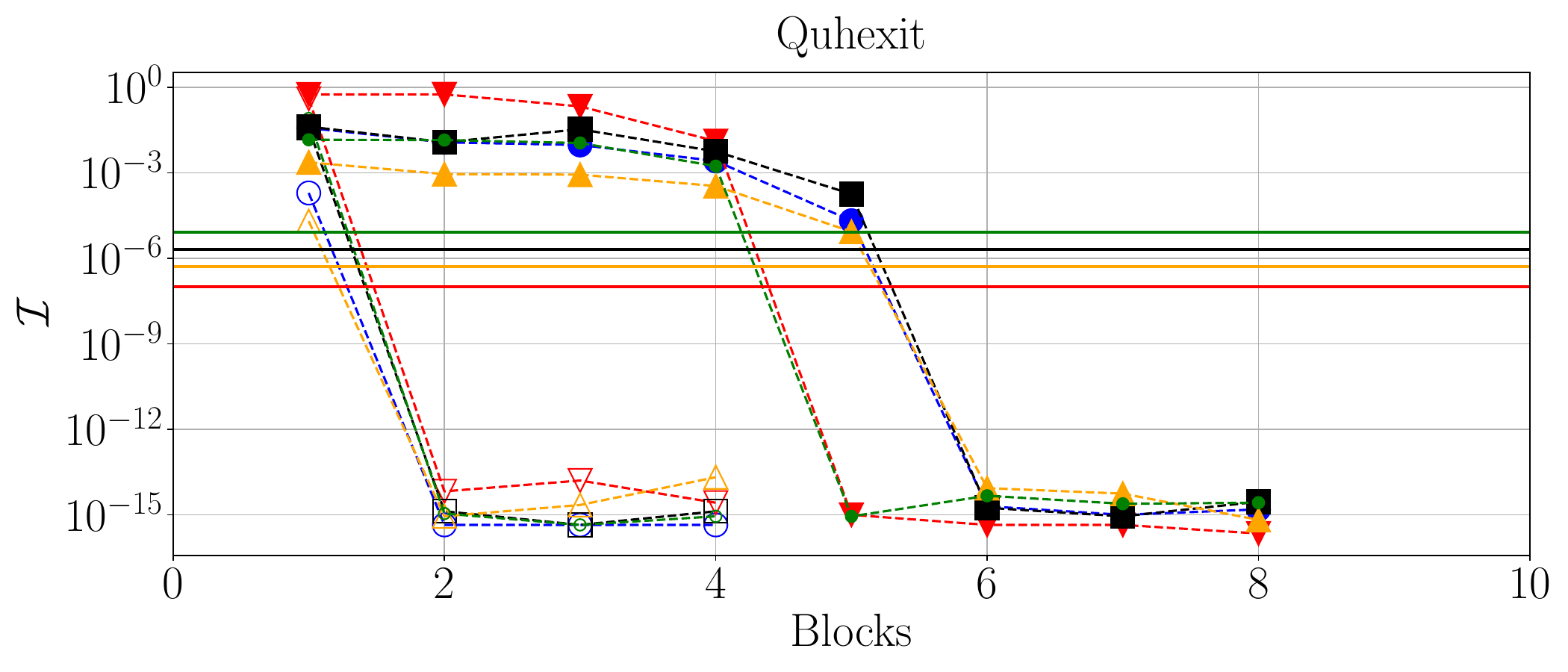}
\includegraphics[width=\linewidth]{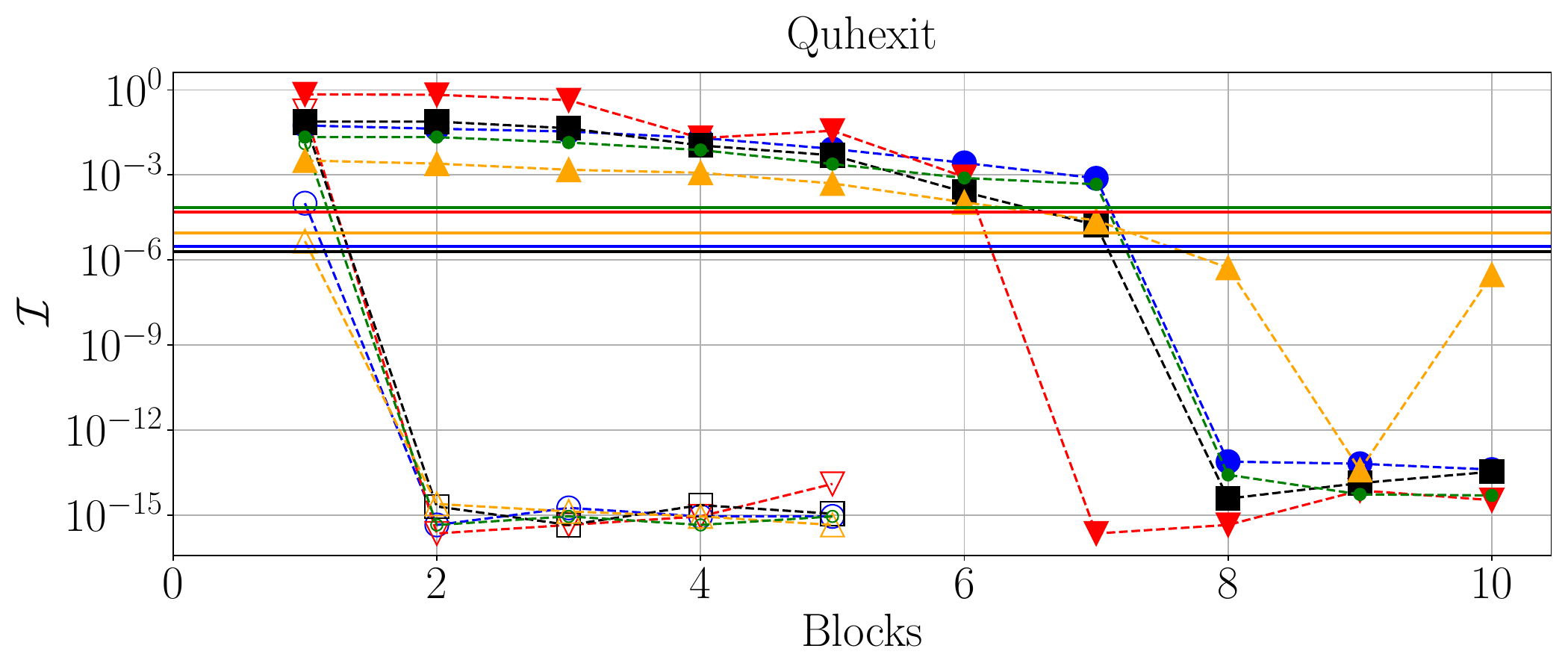}
    \caption{$S+D$ (open points) and $ECD$ (closed points) decompositions of different state preparations for: (top) ququart, (middle) quhexit, (bottom) quoctit. The horizontal lines correspond to the reference values of the PO infidelities.}
    \label{fig:stateprep}
\end{figure}

For each decomposition, $\mathcal{I}$ as a function of $B$ are presented in \fig{fig:stateprep}. From these $\ket{\Psi}$, the $S+D$ decompositions appear independent of qudit size $d$-- always requiring a block size $B_{S+D}=2$ to reach $\mathcal{I}<10^{-15}$.  In contrast reaching $\mathcal{I}<10^{-15}$ requires $B_{\rm ECD}\approx d$. Thus these $\ket{\Psi}$ confirms the expectations of $B_{ECD}> B_{S+D}$ and the scaling from naive parameter counting. Further, while $ECD$ requires more blocks, the potential 2-50$\times$ faster time per block can mitigate this so the ultimate preference depends on experimental hardware. These results can be compared to pulse optimization with total times roughly one block of the ECD gate but different for each qudit: $T_{PO}^4=0.1~\mu s,~T_{PO}^6=0.2 \mu s.~T_{PO}^8=0.5~\mu s$.  Example of the pulses for the quhexit are found in \fig{fig:stateprep}. While these pulses do not start and stop at zero power, imposing smooth ramping was found to have only small effect on $\mathcal{I}$ but greatly increased convergence time.  From this, $\mathcal{I}_{PO}\lesssim 10^{-5}$ has a clear dependence on $d$. 

\begin{table}
\caption{$\mathcal{I}$ from pulse optimization of state preparation on different qudits.}
\centering
\begin{tabular}{cc|ccccc}
                $d$  & T [$\mu$s]   & $X$                   & $R_X$                  & $U_2$                 & $\ket{0} \rightarrow \ket{3}$ & $\ket{0} \rightarrow \ket{H}$ \\ \hline
4&0.1& $2\times 10^{-12}$ & $6 \times 10^{-11}$ & $2\times 10^{-10}$ & $2 \times 10^{-10}$        & $6 \times 10^{-8}$            \\  
6&0.2& $2\times 10^{-6}$ & $5 \times 10^{-7}$ & $8\times 10^{-6}$ & $1 \times 10^{-7}$        & $2 \times 10^{-6}$               \\ 
8&0.5& $3\times 10^{-6}$ & $9 \times 10^{-6}$ & $7\times 10^{-5}$ & $5 \times 10^{-5}$        & $2 \times 10^{-6}$               \\ \hline      
\end{tabular}
\label{tab:state-prep-table}
\end{table}

\begin{figure*}
    \centering
\includegraphics[width=.85\linewidth]{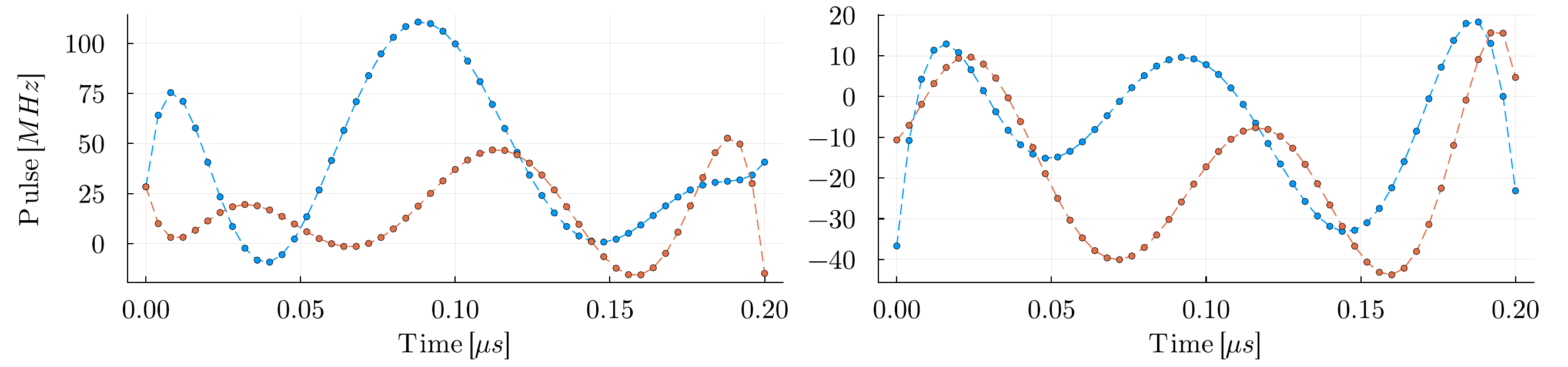}
    \caption{Pulse optimization for the state prep of $H_6\ket{0}=\frac{1}{\sqrt{6}}\sum_i\ket{i}$ for a quhexit}
    \label{fig:stateprep-pulses}
\end{figure*}

\begin{figure}
    \includegraphics[width=\linewidth]{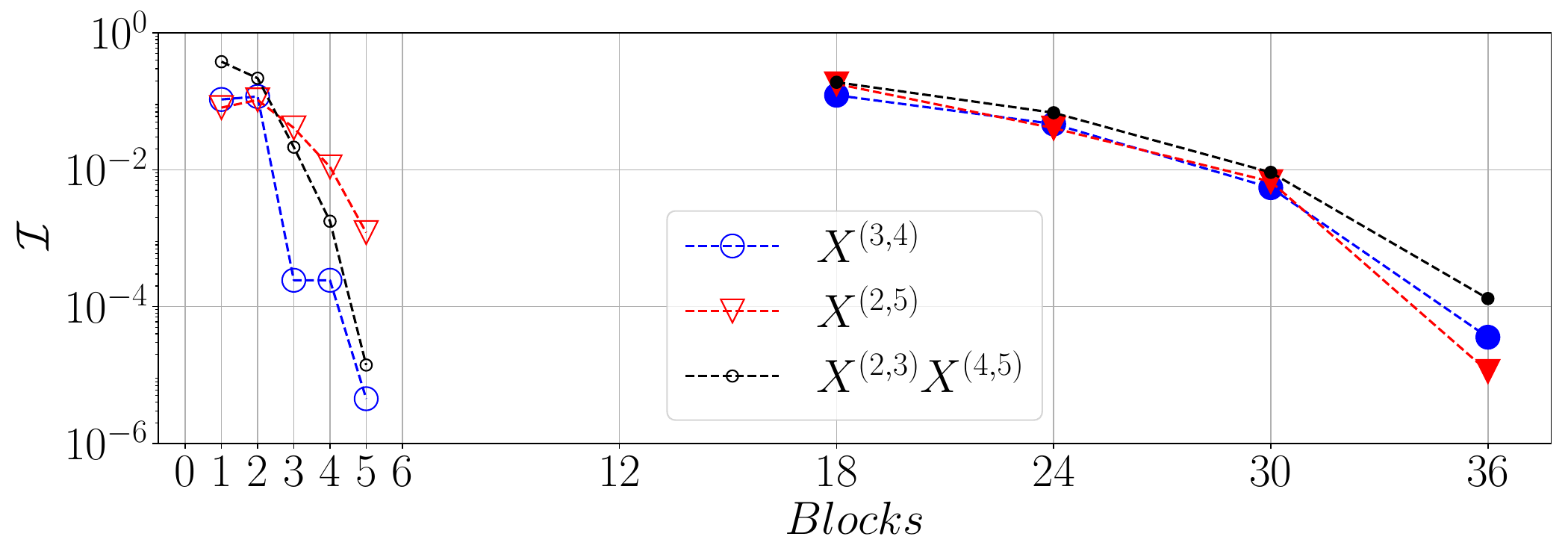}
        \includegraphics[width=\linewidth]{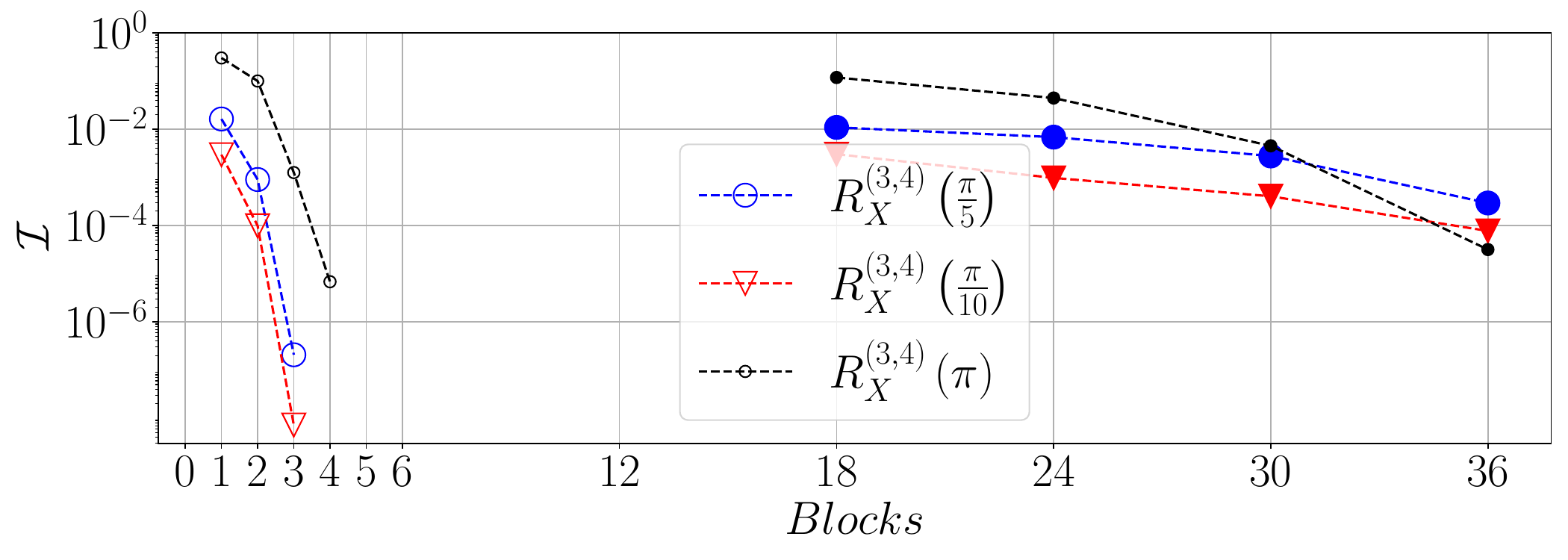}
    \caption{$\mathcal{I}$ vs. blocks for S+D and ECD for quhexit gates.}
    \label{fig:state-dependency}
    \end{figure}

Moving to gate decomposition, we consider the following gates which are prototypical of HEP primitives:
\begin{itemize}
   \item $X^{(3,4)}$, $X^{(2,5)}$, and $X^{(2,3)}X^{(4,5)}$ which can be related to $\mathfrak{U}_{-1}$ and $\mathfrak{U}_{\times}$ 
   \item $R_{X}^{(3,4)}(\theta)$ with $\theta=\pi,\pi/5,\pi/10$ which model $\mathfrak{U}_{\rm Tr}$ and $\mathfrak{U}_{\rm F}$.
\end{itemize}
For these gates, we consider their decomposition on quhexits with $S+D$ and $ECD$ (See \fig{fig:state-dependency}).  In contrast to state preparation, $\mathcal{I}$ depends strongly upon $B$.  We observe little difference in scaling for $X^{(3,4)}$ and $X^{(2,3)}X^{(4,5)}$, but $X^{(2,5)}$ -- with its larger distance between states -- appears to converge more slowly for $S+D$. Notice how $\mathcal{I}$ is not strongly affected by applying multiple $X$-gates -- confirming the S+D observation and suggesting it also holds for ECD decompositions. For the $R_X^{(3,4)}(\theta)$ gates, $\mathcal{I}$ increases with $\theta$ for both decompositions. Further, for fixed $B$, For the small $\theta$ investigated here, $R_X$ has lower $\mathcal{I}$ that $X$ gates.  In all cases $B_{S+D}\ll B_{ECD}$ with the relative factor being $\sim 6=d$ which comports with the scaling from parameter counting.

\begin{figure}
    \centering
    \includegraphics[width=\linewidth]{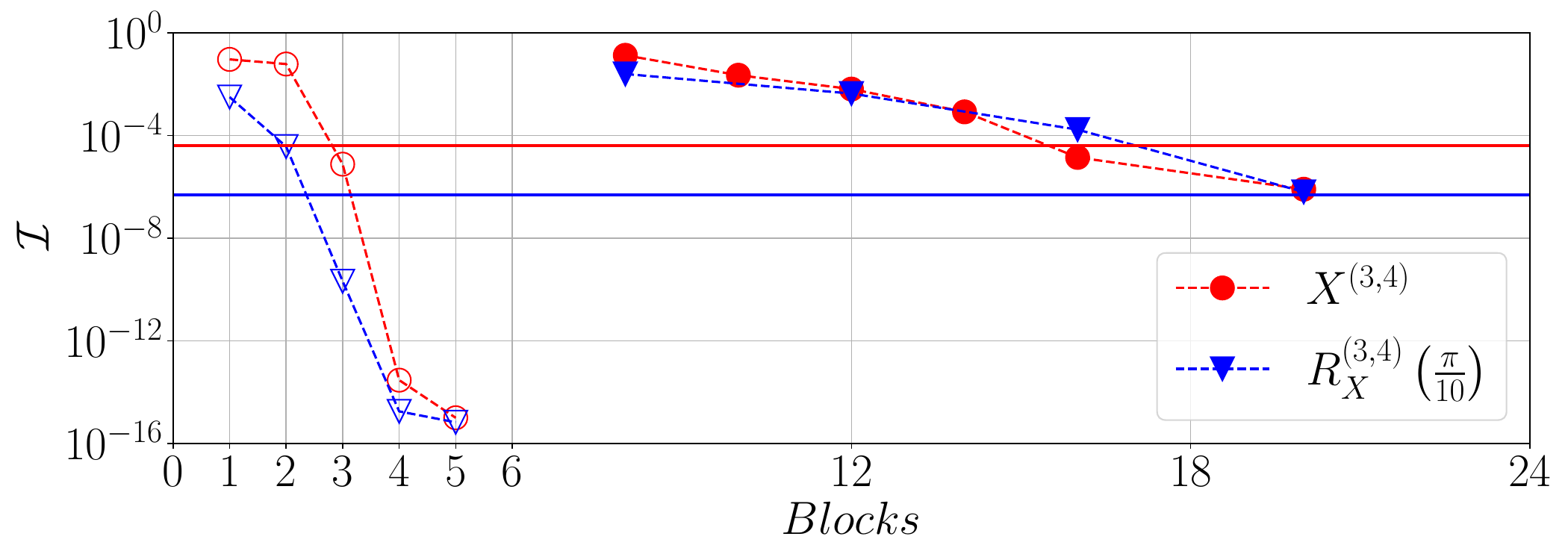}
        \includegraphics[width=\linewidth]{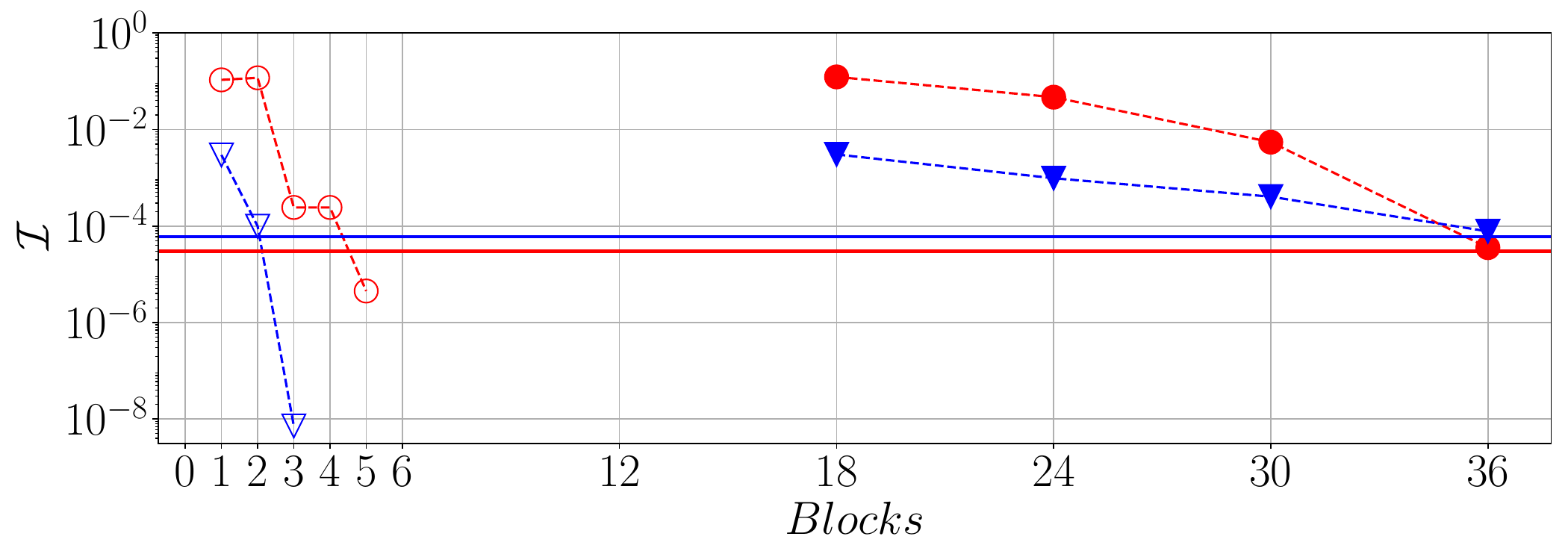}
        \includegraphics[width=\linewidth]{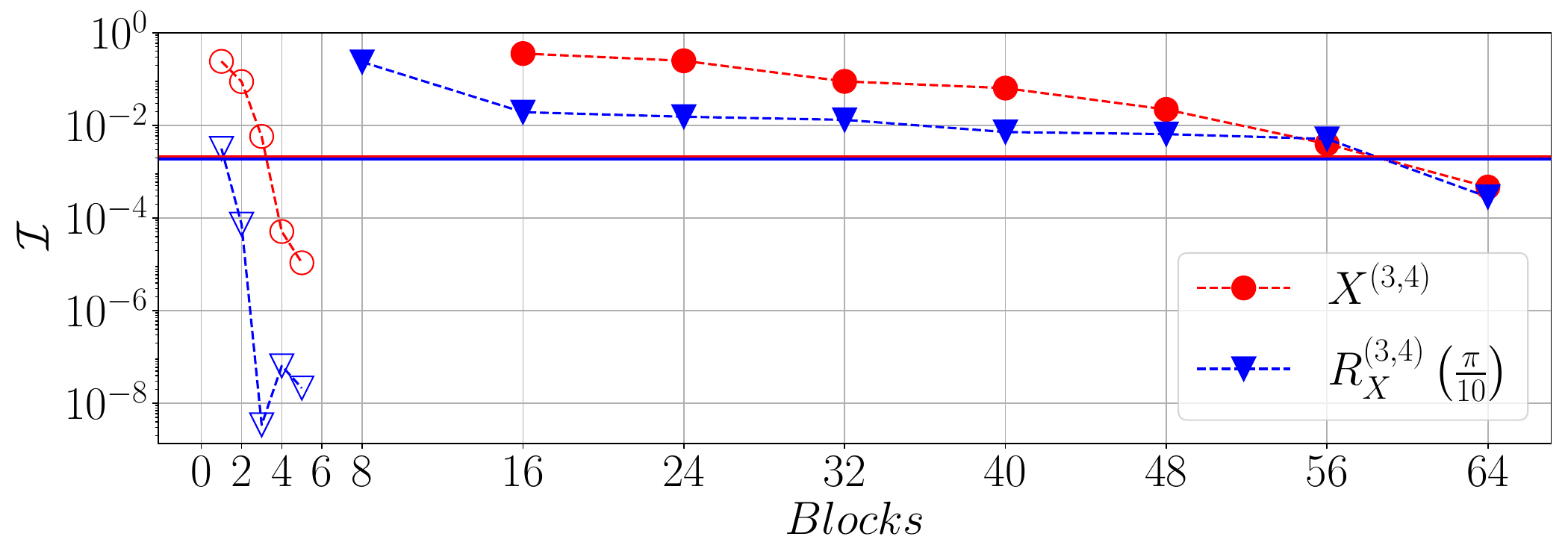}
    \caption{$\mathcal{I}$ vs blocks for $S+D$ and ECD decomposition of an $X^{(3,4)}$ and $R_x^{(3,4)}(\pi/10)$  gate, for ququart, quhexit and quoctit from top to bottom. The horizontal lines correspond to the reference values of the PO infidelities.}
    \label{fig:ququart}
\end{figure}

After looking at the dependence of $\mathcal{I}$ on the two decomposition methods, we perform an analysis of two gates $X^{(3,4)}$ and $R_X^{(3,4)}(\pi/10)$ for $d=4,6,8$ qudits which is presented in \fig{fig:ququart}. If we take as a fiducial $\mathcal{I}=10^{-4}$, we find that nearly constant $B_{S+D}\sim 2-3$ for all $d$, while $B_{ECD}=16,36,64$ for $d=4,6,8$ respectively. This again suggests $B_{S+D}$ scales sublinearly with $d$. In contrast, ECD scales consistently with $B_{ECD}\sim d^2$.
 
Taking these results, one can estimate the break-even dimension $d_b$ when the decompositions require the same amount of time by setting
\begin{equation}
    B_{ECD}T_{ECD}=B_{S+D}T_{S+D}.
\end{equation}
With the  empirically-observed constant $B_{S+D}$ and relative factor of $\sim1$ between the decompositions, one finds
\begin{equation}
    \frac{T_{S+D}}{T_{ECD}}=d_b^2 \implies d_b=\sqrt{\frac{[1,50]~\mu\text{s}}{[0.1,0.5]~\mu\text{s}}}=[2,22],
\end{equation}
if instead, one takes the scaling suggested from parameter counting $B_{S+D}=d$, the range is $d_b=[2,500]$.
% \begin{equation}
%     \frac{T_{S+D}}{T_{ECD}}=d_b\implies d_b=\frac{[1,50]~\mu\text{s}}{[0.1,0.5]~\mu\text{s}}=[2,500].
% \end{equation}
These heuristics suggest while neither gate decomposition is decisively better, hardware-obtainable $T_{S+D}$ and $T_{ECD}$ would resolve it.

\begin{figure}
    \centering
    \includegraphics[width=\linewidth]{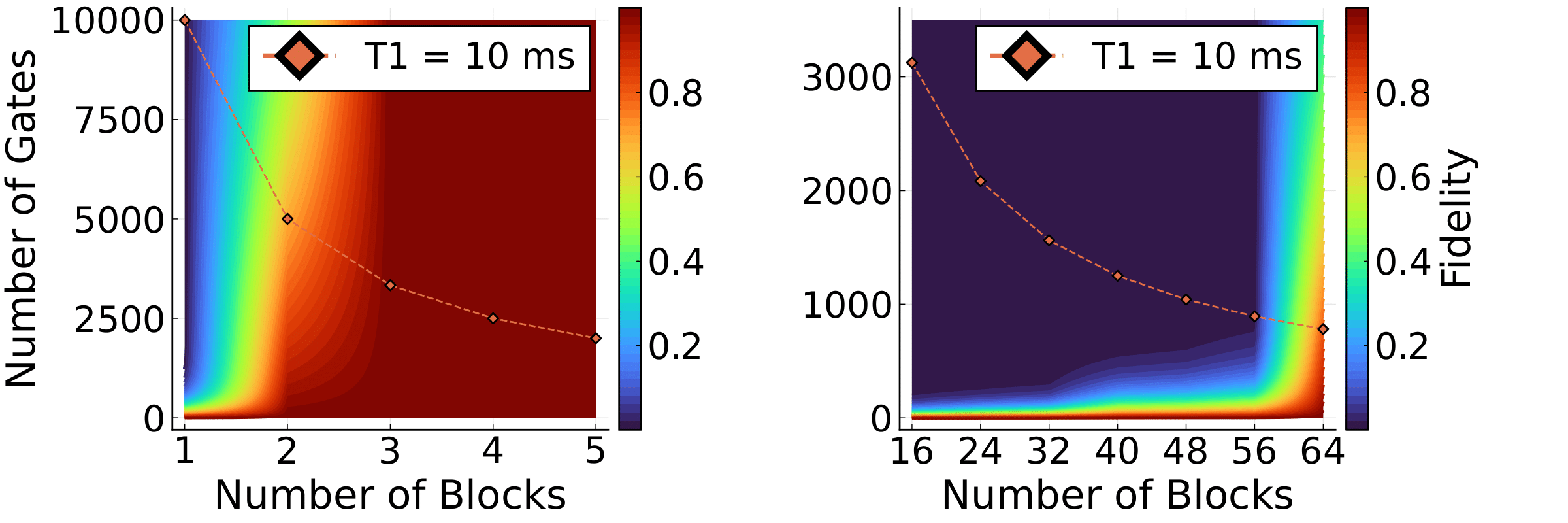}
    \caption{$R_X^{(3,4)}(\pi/10)$ gate decomposition on quoctit using (left) $S+D$ and (right) ECD. The $T_1$ is plotted as a reference for the maximum number of gates given a certain amount of blocks, assuming fixed gate times of $T_{SNAP}=1 \mu s$ and $T_{ECD}=200 ns$.}
    \label{fig:quoctit_contour_rx}
\end{figure}

This logic can also be used to estimate the number of HEP primitive gates could be simulated within $T_1=10$ ms of the qudit with fixed $\mathcal{I}$ versus $B$.  In Fig.~\ref{fig:quoctit_contour_rx} is a fiducial result for the $R_X^{(3,4)}(\pi/10)$ gate decomposition on a quoctit with fixed gate times of $T_{S+D}=1~ \mu s$ and $T_{ECD}=0.2~\mu s$.  Together these suggest such devices could reasonably achieve circuit depth of thousands of gates with total $\mathcal{I}<1\%$. Given the resources estimates for $SU(2)$ in~\cite{Gustafson:2022xdt}, this would be sufficient for quantum utility in toy models.

\begin{figure*}
        \includegraphics[width=1\linewidth]{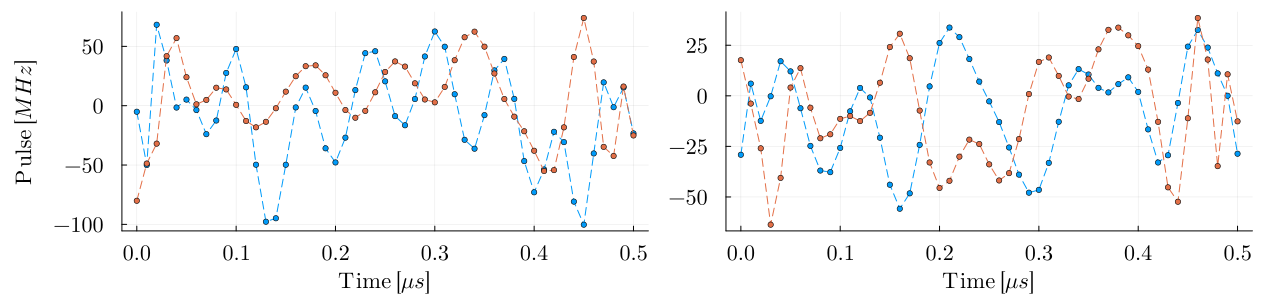}
    \caption{Example $X$ gate pulses for a quhexit, with a control on the pulses every $10 ns$.}
    \label{fig:quhextit-x-pulses}
\end{figure*}

Finally, we consider optimal control. The pulse duration $T_{PO}=0.5~\mu s$ for the ququart and quhexit to be competitive with the single-block gate set decompositions, but had to be extended to 2 $\mu$s for the quoctit to get $\mathcal{I}<10^{-2}$. An example of the resulting pulses is shown in \fig{fig:quhextit-x-pulses}. The lowest infidelity from 10 trials for all three qudit dimensions are shown in \tref{tab:inf-po}. An additional optimization that proved useful for the quoctit was to start with a smaller basis set, truncated at the 32th order. Then every $500$ iterations, new terms were added to the decomposition. The new terms were added in 3 different batches consisting of 8, 5 and 5.

\begin{table}
\caption{$\mathcal{I}_{PO}$ of two gates obtained with pulse optimization for qudits.  For all cases, the time interval was divided into 50 time steps and the best optimization of 10 trials was taken. The final truncation order of the Chebysev series is denoted by $o_C$.}
\centering
\begin{tabular}{ccc|ccc}
                    d & T [$\mu$s] & $o_C$ & $X$                   & $R_x$                  \\ \hline
$4$ & 0.5&18 &$4\times 10^{-5}$ & $5 \times 10^{-7}$               \\ 
$6$ & 0.5&30 & $3\times 10^{-5}$ & $6 \times 10^{-5}$               \\ 
$8$ & 2.0&50 & $2\times 10^{-3}$ & $2 \times 10^{-3}$               \\ \hline
\end{tabular}
\label{tab:inf-po}
\end{table}

For the case of the quhexit, the infidelity from pulse optimization $\mathcal{I}_{PO}\lesssim10^{-5}$ was close to that of a 5-block $\mathcal{I}_{S+D}$ and a 36 block $\mathcal{I}_{\rm ECD}$ which allows us to make a direct comparison in this case of the relative gate time for the $X^{(3,4)}$ and $R_{X}^{(3,4)}(\pi/10)$:
\begin{align}
    T^{6}_{PO}&\approx 0.5~\mu\text{s}\notag\\
    T^{6}_{S+D}&\approx [5,250]~\mu\text{s}\notag\\
    T^{6}_{\rm ECD}&\approx [3.6,18]~\mu\text{s}\notag.
\end{align}
Thus we find that pulse optimization for $X^{(3,4)}$ and $R_{X}^{(3,4)}(\pi/10)$ at fixed infidelity appears to yield at least a factor of 10 in reduction of gate time.  For the quoctit, we can similarly compare to the gate sets, albeit only at a higher $\mathcal{I}\lesssim10^{-3}$ since that was the best obtained from pulse optimization.  For comparable infidelities a 4-block $S+D$ and 64-block ECD were found sufficient,
\begin{align}
    T^{8}_{PO}&\approx 2.0~\mu\text{s}\notag\\
    T^{8}_{S+D}&\approx [4,200]~\mu\text{s}\notag\\
    T^{8}_{\rm ECD}&\approx [6.4,32]~\mu\text{s}\notag.
\end{align}
In this case, the improvement was a more modest factor for 2, which may indicate a diminishing in advantage at higher dimension or a reflection of the difficulty of classical optimization at high dimension.

\begin{figure}
    \centering
        \includegraphics[width=\linewidth]{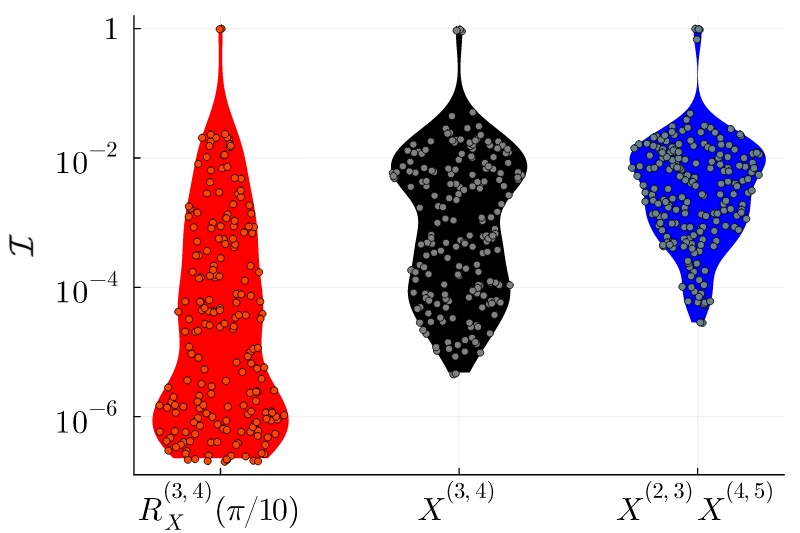}
    \caption{Distribution of infidelities for 5-block $S+D$ decomposition for different quhexit gates starting from the same initial parameters.}
    \label{fig:boxplot-comparison}
\end{figure}

While these results demonstrate it is possible to decompose gates of interest to a universal set, the decomposition generically has poor scaling $\mathcal{O}(d^n)$ with $n\geq 3$~\cite{Fosel:2020oyj,Job:2023huu,Ogunkoya:2023cfg} that without mitigation will become prohibitive for large $d$. It is therefore worth investigating the relations between optimized parameters to seek heuristics to guide the optimization. Since $S+D$ requires fewer blocks than ECD, we will restrict our analysis to this decomposition and study the 5-block versions of $X^{(3,4)}$, $X^{(2,3)}X^{(4,5)}$, $R_X^{(3,4)}(\pi/10)$ gates acting on a quhexit. The optimization is performed 200 times for each gate. The distributions of $\mathcal{I}$ are shown in \fig{fig:boxplot-comparison}, where one observes a clear ordering of $\mathcal{I}$ from smallest to largest of $R_X^{(3,4)}(\pi/10)$, $X^{(3,4)}$, $X^{(2,3)}X^{(4,5)}$.  This supports the intuition that the distance from $\mathbb{1}$ is a proxy for difficulty.

\begin{figure}
    \centering
    \includegraphics[width=\linewidth]{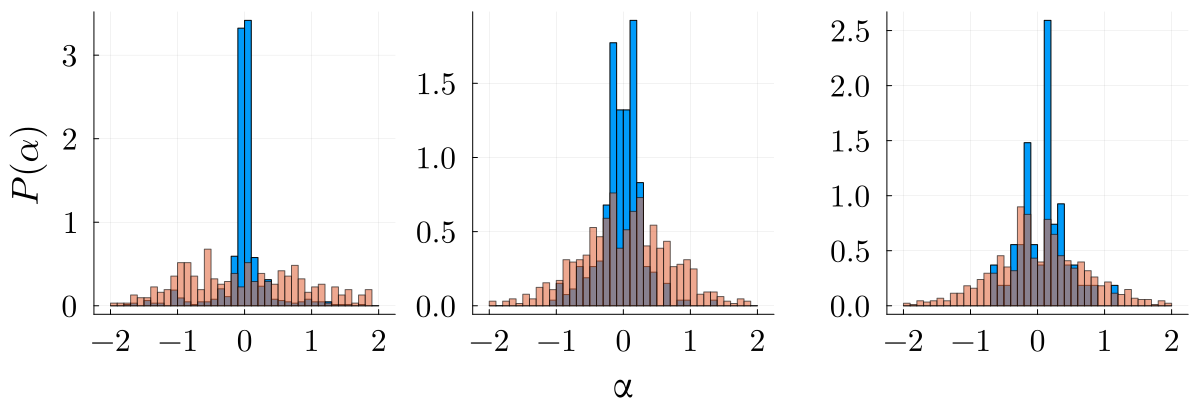}
    \caption{Distribution of $\alpha_k$ of quhexit optimizations that gave $\mathcal{I}> 10^{-4}$ (orange) and $\mathcal{I}\le 10^{-4}$ (blue). From left to right for $R_X^{(3,4)}(\pi/10)$, $X^{(3,4)}$ and $X^{(2,3)}X^{(4,5)}$}
    \label{fig:disp-distribution}
\end{figure}

Probing further, one can study the distribution of $S+D$ parameters. The distribution of the SNAP $\theta_k$ appear to be uniformly distributed regardless of $\mathcal{I}$.
In contrast, the displacement parameters clearly distinguish between good and poor optimizations as seen in \fig{fig:disp-distribution}. For $\mathcal{I}\le 10^{-4}$ where $\alpha$ cluster near 0, compared to optimizations where $\mathcal{I}\ge 10^{-4}$ that have broader distributions.  This can be used to reduce the search parameter space.  

\begin{figure}
    \centering
        \includegraphics[width=0.49\linewidth]{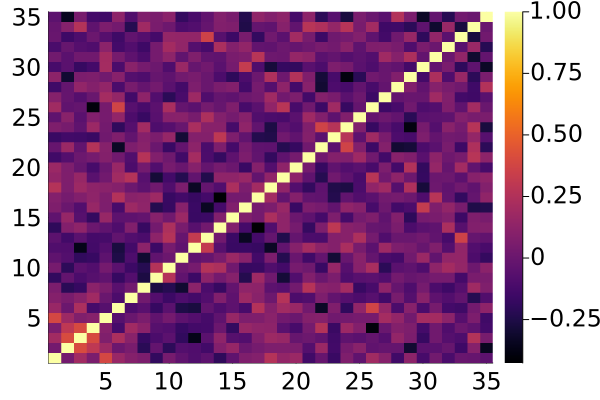}
                \includegraphics[width=0.49\linewidth]{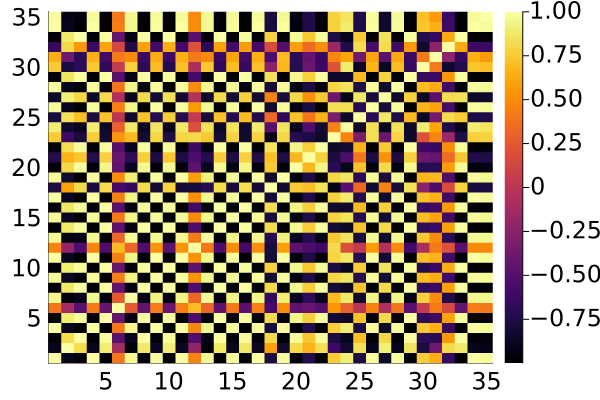}
    \caption{Correlation matrix across parameters for $S+D$ decompositions of $X^{(3,4)}$ gate for a 5-block quhexit optimizations with (left) $\mathcal{I}\le 10^{-5}$ and (right) $\mathcal{I}> 10^{-5}$.}
    \label{fig:correlation-matrices-good}
\end{figure}

\begin{figure}
    \centering
    \includegraphics[width=\linewidth]{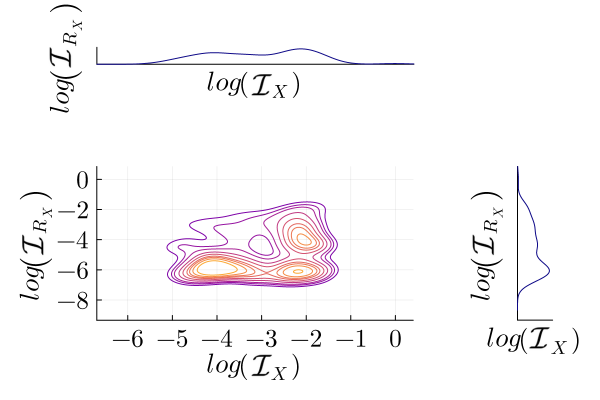}
    \caption{Contours of the infidelities reached from same initial parameters for a $R_X$ and a $X$ gate.}
    \label{fig:2d-hist-snap}
\end{figure}

 Another source for accelerating the optimization would use  correlations between the optimal parameters. This is further motivated by empirical observations that for optimal $B_{S+D}=1$ decompositions $\alpha_1=-\alpha_0$ of many gates. We searched for correlations across optimizations of  $X^{(3,4)}$. \fig{fig:correlation-matrices-good} shows the correlation matrix corresponding to high and low infidelities. The correlation between two decomposition parameters $X,Y\in\{\alpha_k,\theta_k\}$ were computed using \begin{equation*}
     \rho(X, Y) = \frac{1}{\sigma_X \sigma_Y(n-1)}\sum_{i=1}^{n} (X_i - \bar{X})(Y_i - \bar{Y}),
 \end{equation*}
 where $\sigma_X$ and $\bar{X}$ are the standard deviation and mean of $X$. From the analysis of these correlations, we found different distributions when comparing low and high infidelity decompositions, when $\mathcal{I}>10^{-5}$ the parameters are highly symmetric and with large correlation (or anticorrelation). In contrast, for $\mathcal{I}\leq10^{-5}$ the parameters display decreasing correlation with gates farther away in the decomposition i.e. gates in the same block are more correlated.  Additionally, there is some clear, weak anti-correlation structure suggesting new ansatze for decomposition.  Moreover, we found that the same initial conditions gave either good results for two different gates, or bad results for both.  This is presented in Fig.~\ref{fig:2d-hist-snap} and suggests that there might be only a subspace of the parameters. However, this hypothesis requires more analytical work should be explored in the future. 

A final consideration is the robustness to quantum noise. As a first step toward understand this, we studied how $\mathcal{I}$ degrades when the parameters of the gate sets and pulse optimization were perturbed by a Gaussian noise.  This basic noise model could be taken to approximate the stochastic fluctuations in the pulses driving the cavity and qubit. In this model the parameters are varied:
\begin{equation}
\label{eq:stab-analysis}
    \alpha\rightarrow\alpha(x) =\alpha + \frac{1}{\beta |\alpha|\sqrt{2 \pi} }\exp^{-\frac{1}{2}\left(\frac{x}{\beta |\alpha|}\right)^2},
\end{equation}
where $\alpha$ is the parameter of the decomposition, and $\beta$ is the square-root of the variance. 

\begin{figure}
    \centering
    \includegraphics[width=\linewidth]{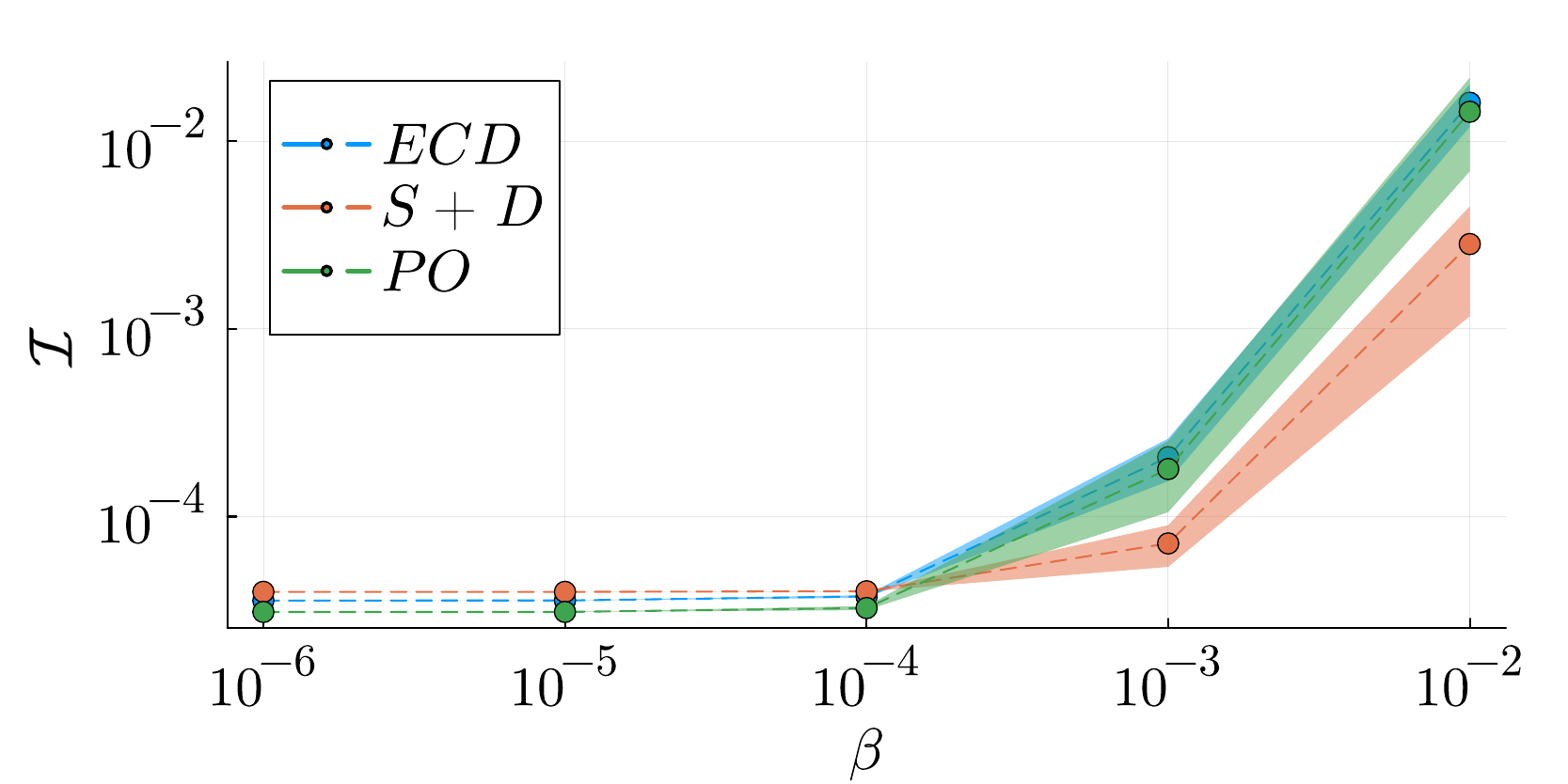}
    \caption{Stability analysis for all the decompositions for a $X$ gate on a quhexit.}
    \label{fig:stability-analysis}
\end{figure}

Taking the noiseless $X^{(3,4)}$ gates on the quhexit with $\mathcal{I} \approx 10^{-5}$: 5-block $S+D$ with $\mathcal{I} = 3.97 \times 10^{-5}$, 36-block ECD with  $\mathcal{I} = 3.57 \times 10^{-5}$, and 30th order pulse optimization for 0.5~$\mu$s with $\mathcal{I} = 3.10 \times 10^{-5}$;  50 random sample perturbations for each $\beta$ are computed. The plot showing the results is \fig{fig:stability-analysis}. One observes that for $\beta<10^{-4}$, the effect of noise is negligible.  At larger $\beta$, pulse optimization and $ECD$ are found to demonstrate similar sized infidelities from the noise, while $S+D$ is found to have $\mathcal{I}$ about half the size of the other two.

\section{Conclusion}
\label{sec:conc}
In this work, we compared the three methods for implementing prototypical primitive gates for simulating lattice gauge theories on a 3D SRF qudit architecture: Snap \& Displacement gates, ECD \& single-qubit rotations, and optimal pulse control. We observed that lattice gauge theory primitives have common properties that distinguish them from general $SU(d)$ circuits,  and thus the choice of decomposition could greatly change the algorithmic reach of devices. We found evidence that for subsets of permutation gates the number of SNAP and Displacement blocks required to approximate them scales sublinearly with qudit dimension -- in contrast to the general case of linear scaling --, while the ECD block depth quadratically. In all cases, we found that decomposing a sequence of $X^{(a,b)}$ gates rather than individual reduces the total number of blocks required and consequently the total time for fixed infidelity.  With the gate sets, we found that $\mathcal{I}<10^{-5}$ is possible for $B_{S+D}=d$ and $B_{ECD}=d^2$. In addition, pulse decomposition with modest classical resources can outperform in run time for $d<8$, while becoming comparable to gate decomposition for quoctits. Beyond run time, the $S+D$ decomposition was found to be more robust to Gaussian perturbations.

 From the analysis of the $S+D$ decomposition, we found that good gate decompositions can be distinguished by symmetries, correlations, and the distribution of parameters. Further, we found correlations between initial conditions, suggesting a subspace useful as initial conditions. 

 From this initial work, we found that algorithmic advantage and hardware-specific gate times can be used to decisively choose between different native gate sets on qudit devices for the restricted class of gates needed for high energy physics. In the future, more in-depth investigation of the scaling of the decompositions should be performed. Further, it would be of invaluable help to try to implement some of the decompositions on the actual hardware, in order to better understand and develop a noise model and how the infidelities are affected.  

\begin{acknowledgements}
The authors would like to thank Tanay Roy, Nicholas Bornman, and Judah Unmuth-Yockey for helpful discussions. This material is based on work supported by the U.S. Department of Energy, Office of Science, National Quantum Information Science Research Centers, Superconducting Quantum Materials and Systems Center (SQMS) under contract number DE-AC02-07CH11359. Fermilab is operated by Fermi Research Alliance, LLC under contract number DE-AC02-07CH11359 with the United States Department of Energy.
\end{acknowledgements}

\bibliography{wise}

% \appendix

% \section{Best optimization runs for multiple qudit sizes and different gates}
% In this section, we provide the best optimization runs obtained for all the gates and all the qudit sizes tested in this work. The focus for the ECD decomposition and SNAP and displacement decomposition was on the best possible value obtained, without taking into consideration the number of blocks required to get that value. 

% \begin{table}[h]
% \caption{Best optimization runs for gate decomposition for different qudit dimensions and different decomposition approaches.}
% \begin{tabular}{ccccc}
%                                 d & Decomp. & Ququart               & Quhextit              & Quoctit               \\ \hline
% $X$ gate &S\&D   & $\approx 0$           & $4.6 \times 10^{-6}$ & $1.5 \times 10^{-5}$ \\
% $R_x$ gate& S\&D & $\approx 0$           & $7.3 \times 10^{-9}$ & $2.6 \times 10^{-7}$ \\
% $X$ gate& ECD                     & $8.2 \times 10^{-7}$ & $3.6 \times 10^{-5}$ & $4.7 \times 10^{-4}$ \\
% $R_x$ gate& ECD                   & $6.4 \times 10^{-7}$ & $7.8 \times 10^{-5}$ & $2.9 \times 10^{-4}$ \\
% $X$ gate &PO      & $3.3 \times 10^{-5}$ & $3.1 \times 10^{-5}$ & $1.9 \times 10^{-3}$ \\
% $R_x$ gate& PO    & $4.6 \times 10^{-7}$ & $5.8 \times 10^{-5}$ & $2.1 \times 10^{-3}$ \\ \hline
% \end{tabular}
% \label{table:summary-table}
% \end{table}

\end{document}